\documentclass[a4paper,11pt]{article}
\pdfoutput=1 
\usepackage{jheppub} 
                     
\usepackage{booktabs,makecell, bm}
\usepackage[font=small]{caption}
\usepackage{framed}
\usepackage{hyperref}
\usepackage[export]{adjustbox}
\usepackage{amsmath}
\DeclareMathOperator{\Tr}{Tr}
\usepackage{physics}
\usepackage{subcaption}

\newcommand{\C}[1]{\mathcal{#1}}
\newcommand{\B}[1]{\overline{#1}}
\newcommand{\T}[1]{\Tilde{#1}}

\title{Holographic Tests for Giant Graviton Expansion}

\author[a]{Seunggyu Kim}
\author[b]{and Eunwoo Lee}

\affiliation[a]{Department of Physics, Korea Advanced Institute of Science and Technology\\291 Daehak-ro, Yuseong-gu, Daejeon 34141, Republic of Korea.}
\affiliation[b]{Department of Physics and Astronomy \& Center for
Theoretical Physics,\\
Seoul National University, Seoul 08826, Korea.}

\emailAdd{sgkim01@kaist.ac.kr}
\emailAdd{eunwoo42@snu.ac.kr}

\preprint{SNUTP23-003}

\abstract{
It has been proposed that the superconformal index admits a novel reformulation, called giant graviton expansion. In this paper, we investigate the properties of dual $AdS_5$ black holes using the giant graviton expansion framework.
First, we compute the entropy of black holes in $AdS_5\times S^5$ with fixed charges through a large $N$ saddle point analysis on the giant graviton index and further extremize it in the wrapping number. 
We identify a specific regime of fugacities where our saddle point analysis is valid. It turns out that this condition ensures the absence of closed-time-like curves and the stability of dual black hole solutions with equal charges.
In addition, the giant graviton expansion of the index provides insights into how small black holes in AdS can be interpreted as bound states of branes. 
We extend our study to include the giant graviton expansion with the insertion of a half-BPS surface defect in $\mathcal{N}=4$ SYM with a $U(N)$ gauge group. Finally, we test the giant graviton expansion in various holographic theories whose dual geometries are $AdS_5\times S^5/\mathbb{Z}_k$ and $AdS_5\times SE_5$.}

\begin{document} 
\maketitle

\section{Introduction}
Following the pioneering work of Strominger and Vafa \cite{Strominger:1996sh}, there has been extensive progress in understanding the microscopic origin of the entropy of black holes via the holographic principle. In particular, the Bekenstein-Hawking entropy of the BPS black holes in $AdS_5\times S^5$ \cite{Gutowski:2004ez,Gutowski:2004yv,Chong:2005hr,Kunduri:2006ek} was derived microscopically by computing the superconformal index, which counts the degeneracy of gauge-invariant BPS states of the $\C{N}=4$ SYM living in the boundary \cite{Cabo-Bizet:2018ehj, Choi:2018hmj, Benini:2018ywd}. Since then, various developments have been made using different methods, such as Cardy-like limits \cite{Choi:2018hmj, Choi:2018vbz, Honda:2019cio, ArabiArdehali:2019tdm, Kim:2019yrz, Cabo-Bizet:2019osg, Amariti:2019mgp, Goldstein:2020yvj, Jejjala:2021hlt}, and Bethe-Ansatz type formula \cite{Closset:2017bse, Benini:2018mlo} for the large $N$ index \cite{Benini:2018ywd, Lanir:2019abx, GonzalezLezcano:2019nca, Benini:2020gjh, Mamroud:2022msu, Aharony:2024ntg} and a particular extension of the index integral \cite{Cabo-Bizet:2019eaf, Cabo-Bizet:2020nkr}. Furthermore, recently, it was found that the matrix integral of the index admits a genuine saddle in large $N$, called parallelogram ansatz, which accounts for the $O(N^2)$ free energy of the dual black holes \cite{Choi:2021rxi}. This large $N$ saddle is universal in that it covers a large class of known holographic Lagrangian gauge theories with a suitable large $N$ limit \cite{Choi:2023tiq}.

Meanwhile, the giant graviton (GG) expansion of the index has been recently conjectured \cite{Gaiotto:2021xce, Imamura:2021ytr} (see also for related works \cite{Lee:2022vig, Murthy:2022ien, Liu:2022olj, Lin:2022gbu, Eniceicu:2023uvd, Lee:2023iil, Eleftheriou:2023jxr}). This reformulation enumerates D/M-branes and stringy excitations from the gravity (dual string theory) perspective. It captures the finite $N$ corrections to the supergravity index, which can be interpreted as the effect of the BPS configurations of D/M-branes wrapping cycles of the internal manifold, whose charges are of order $O(N)$. These branes are usually called giant gravitons \cite{McGreevy:2000cw,Grisaru:2000zn,Hashimoto:2000zp,Balasubramanian:2001nh}.

Formally, the GG-expansion of the index is given by
\begin{align}\label{eqn: GGE}
    \frac{\C{I}_N}{\C{I}_{\infty}}=\sum_{m_1,\cdots, m_d}x_1^{m_1 N}\cdots x_d^{m_d N}\;\T{\C{I}}_{(m_1,\cdots, m_d)}\ .
\end{align}
where $\C{I}_N$ is the index of the boundary theory with a finite gauge group rank $N$ and $\C{I}_\infty$ corresponds to the index of the low-lying supergravity modes in AdS with no $N$-dependence. $x_1,\cdots,x_d$ are $U(1)$ fugacities associated to the giant gravitons. On the right-hand side, the \textit{giant graviton index} $\T{\C{I}}_{(m_1,\cdots, m_d)}$ is that of the worldvolume theory realized on $(m_1,\cdots,m_d)$ stack of D/M-branes wrapped around $d$ different supersymmetric (and possibly topological) cycles. It reflects the finite $N$ effects of BPS configurations of D/M-branes, including open string excitations on their intersections.
The terms $x_1^{m_1N}\cdots x_d^{m_dN}$ correspond to the classical configurations from these wrapped D/M-branes. The GG-expansion was initially considered to account for the finite $N$ corrections to the index and confirmed by analyzing the modes on D/M-branes at lower rank \cite{Imamura:2021dya, Fujiwara:2021xgu, Arai:2020uwd, Arai:2020qaj, Arai:2019aou, Arai:2019wgv, Arai:2019xmp}. This raises an interesting question --- can the GG-expansion of the index in large $N$ reproduce the entropy of the dual black holes with $O(N^2)$ charges?

In \cite{Choi:2022ovw}, the authors first attempted to demonstrate that the GG-expansion of the index for the $\C{N}=4$ SYM with $U(N)$ gauge group can reproduce the entropy of the dual black hole in $AdS_5\times S^5$. The key idea employed in \cite{Choi:2022ovw} was to perform a large $N$ analysis of the giant graviton index $\T{\C{I}}_{(m_1\cdots,m_d)}$ using saddle point approximation, and then to extremize the GG-expansion \eqref{eqn: GGE} with respect to `real' wrapping numbers $(m_1,\cdots,m_d)\sim O(N)$. 
In the small black hole limit $Q\ll N^2$, one can reproduce the entropy of the dual black hole at the leading order in $N$. Physically, this implies that the small black hole, whose size is much smaller than that of $AdS_5$, can be qualitatively described as a bound state of D-branes. Consequently, it provides a new way of counting the microstates of small black holes in terms of strings and D/M-branes in the bulk side, which is more reminiscent of the seminal work done by Strominger and Vafa \cite{Strominger:1996sh}.\footnote{Some heuristic explanations of the small black holes as a bound state of D-branes were already investigated in \cite{Kinney:2005ej}. In Section~\ref{sec: Black hole with equal charges}, we revisit this direction within the framework of GG-expansion.} 

On the other hand, proceeding with the same approach to extract the entropy of the dual large black hole $Q\gg N^2$, one finds the apparent overestimation of the entropy at leading order in $N$ \cite{Choi:2022ovw}. This discrepancy is naively expected due to the Hagedorn behavior at high energy, which originates from the brane and string configurations in the GG-expansion formula. It renders the partition function divergent and ill-defined for large charges \cite{Hagedorn:1965st,Atick:1988si,Choi:2022ovw} unless fine-tuned cancellations occur between giant graviton indices $\T{\C{I}}_{(m_1,\cdots,m_d)}$.
To address this issue, we follow a cancellation procedure suggested in \cite{Choi:2022ovw}, which involves extremizing the giant graviton index with respect to `complex' wrapping numbers. By applying this principle, we demonstrate the reproduction of the dual black hole entropy at any size across various examples of AdS/CFT.\footnote{In our extremization procedure in general complex domain, we approximate the summation over discrete wrapping numbers $(m_1,\cdots,m_d)$ by integration. This approach is generally not allowed, as it requires fine-tuned cancellation between giant graviton indices for the approximation to be justified. Recently, it has been shown that the cancellation at large charge regime can occur through localization arguments in \cite{Beccaria:2023hip}. We believe that such cancellation mechanisms must also occur for any finite size.} By doing so, we demonstrate that careful consideration of the $O(N)$ and $O(1)$ degrees of freedom accurately captures the degeneracy of the dual black hole saddles at $O(N^2)$.

\begin{figure}
    \centering
    \includegraphics[scale=0.3]{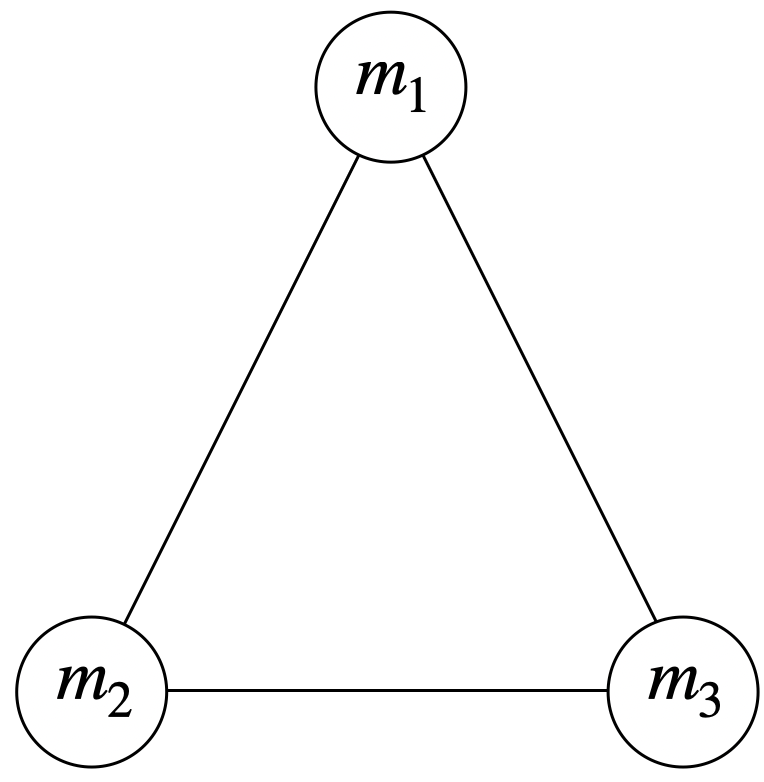}
    \caption{The quiver gauge theory realized on D3-branes with wrapping number $(m_1,m_2,m_3)$, where the bifundamental hypermultiplet represents open strings attached on their intersections.}
    \label{fig: N=4 GGE}
\end{figure}

In general, the GG-expansion \eqref{eqn: GGE} is given by a multiple-sum expansion, which makes the large $N$ analysis difficult. In particular, constructing a large $N$ saddle point of the $\T{\C{I}}_{(m_1,\cdots,m_d)}$ is technically challenging. 
For $\C{N}=4$ $U(N)$ SYM studied in \cite{Choi:2022ovw}, for instance, the GG-expansion is given by a triple-sum, as depicted in the quiver diagram in Figure~\ref{fig: N=4 GGE}. In this case, one cannot find a saddle that solves the large $N$ matrix integral of the giant graviton index $\T{\C{I}}_{(m_1,m_2,m_3)}$, unless the three chemical potentials for the $U(1)^3\subset SO(6)$ charges are equal \cite{Choi:2022ovw}.

In this paper, we improve this result by considering a non-equal chemical potential regime, which corresponds to black holes with non-equal charges. This can be achieved by the reduction of a multiple-sum of GG-expansion to a \textit{reduced-sum}, using analytic properties of the giant graviton index $\T{\C{I}}_{(m_1,\cdots,m_d)}$ \cite{Imamura:2022aua, Fujiwara:2023bdc, Gaiotto:2021xce}. Physically, this means that one can ignore some contributions from giant gravitons and open strings on their intersections. It drastically reduces the complexity of analyzing the large $N$ behavior of the GG-expansion to reproduce the entropy of the dual black hole. For instance, the GG-expansion of $\C{N}=4$ $U(N)$ SYM is given by one parameter expansion, which we call \textit{simple-sum}. Furthermore, using the simple-sum for the $\C{N}=4$ $U(N)$ SYM with equal chemical potentials, we find that the conditions for admitting the ansatz called parallelogram ansatz \cite{Choi:2021rxi} on wrapped D-branes can be identified with not only the closed timelike curve (CTC) bound \cite{Chong:2005hr} but also the instability \cite{Aharony:2021zkr} of the dual black hole solution.

We also study a half-BPS surface defect in $\C{N}=4$ $U(N)$ SYM \cite{Gukov:2006jk, Nakayama:2011pa} within the (simple-sum) GG-expansion framework. The insertion of the surface defect corresponds to considering a probing D3-brane in $AdS_5\times S^5$ background. In large $N$ limit, the surface defect serves as an order parameter for the deconfinement transition \cite{Chen:2023lzq, Cabo-Bizet:2023ejm} whose expectation value is of order $O(N)$ in the black hole background and $O(1)$ in the AdS vacuum. 
We confirm that the defect index can be rewritten in terms of GG-expansion and shows the same features as an order parameter. 
Also, although the surface defect D3-brane and the giant graviton D3-brane are different objects, they can be incorporated in the same kinematic configurations in the GG-expansion framework.

Finally, we explore other holographic theories, whose dual geometries take the form of $AdS_5\times\C{X}$, such as orbifolds, conifolds, and Sasaki-Einstein manifolds, to test the validity of the GG-expansion. In general, the sum over $d$ variables inherent in \eqref{eqn: GGE} simplifies to that of $d-2$ variables \cite{Fujiwara:2023bdc} at most. Once again, using the parallelogram ansatz, we precisely reproduce the entropy of the dual black hole.

The paper is organized as follows: in Section~\ref{sec: Black hole entropy from the boundary theory}, we review how the entropy of the dual black hole in $AdS_5\times S^5$ can be obtained by a saddle point approximation of the index for $\C{N}=4$ $U(N)$ SYM. In Section~\ref{sec: Black hole entropy from the giant graviton expansion}, we reproduce the entropy of the dual black hole using the simple-sum formula of the GG-expansion. In Section~\ref{sec: Black hole with equal charges}, we focus on the black hole with equal charges, where one can find several interesting features of the dual black holes within the GG-expansion framework. For instance, we find an effective $2d$ theory to describe the small black hole. In Section~\ref{sec: Other examples}, we apply the GG-expansion to other holographic theories and see how it reproduces the free energy of the dual black hole in a more complicated background geometry. We conclude in Section~\ref{sec: conclusion} with discussions. In Appendix~\ref{sec: appendix A}, we review the parallelogram ansatz. In Appendix~\ref{sec: appendix B}, we derive the matrix integral of the index on D3-branes wrapped around $S^3/\mathbb{Z}_k$.

\section{Giant graviton expansion in $\mathcal{N}=4$ SYM}
In this section, we begin by reviewing the index of $\C{N}=4$ SYM with $U(N)$ gauge group. Next, we introduce the GG-expansion and explain how it can be simplified further by choosing expansion variables appropriately. Finally, we show the simple-sum formula of GG-expansion precisely reproduces the dual black hole entropy in $AdS_5\times S^5$.

\subsection{Black hole entropy from the boundary theory}\label{sec: Black hole entropy from the boundary theory}
Let us first discuss how the entropy of the dual black hole in $AdS_5$ can be computed from the index using a genuine large $N$ saddle point approximation. The index is defined by \cite{Kinney:2005ej, Romelsberger:2005eg}
\begin{align}\label{eqn: N=4 SYM (1)}
    \C{I}_{U(N)}=\Tr\left[(-1)^F p^{J_1}q^{J_2}x^{Q_1}y^{Q_2}z^{Q_3}\right]\ ,
\end{align}
with a constraint $pq=xyz$. It counts $1/16$-BPS states (or, equivalently, local BPS operators via operator-state correspondence) on $S^3$ annihilated by a pair of complex supercharges $\mathcal{Q},\mathcal{Q}^\dagger$, satisfying the relation:
\begin{equation}\label{eqn: supercharge}
    \{\mathcal{Q},\mathcal{Q}^\dagger\}=E-J_1-J_2-Q_1-Q_2-Q_3=0\ ,
\end{equation}
where $E$, $J_1$, $J_2$, $Q_1$, $Q_2$, and $Q_3$ are Cartan generators of the $SO(2)\times SO(4)\times SU(4)_R$, which are bosonic subgroups of the $PSU(2,2|4)$ superconformal symmetry. Note that the supercharge $\C{Q}$ carries the quantum number 
\begin{equation}
    (J_1,J_2,Q_1,Q_2,Q_3)=\left(-\frac{1}{2},-\frac{1}{2},\frac{1}{2},\frac{1}{2},\frac{1}{2}\right)\ .
\end{equation}
In order to analyze the index in the large $N$ limit, it is useful to express it as a unitary matrix integral, which can be derived through the localization technique in the path-integral formalism
\begin{equation}\label{eqn: N=4 SYM (2)}
    \C{I}_{U(N)}=\int_{U(N)}dg\;\textrm{PE}\left[f_v(p,q,x,y,z)\chi_{adj}^{U(N)}(g)\right]\ ,
\end{equation}
where `$\textrm{PE}$' stands for the plethystic exponential, $\int_{U(N)}dg$ is the gauge group Haar measure, $\chi_{adj}^{U(N)}(g)=\Tr U\Tr U^\dagger$ is the adjoint character, and $f_{v}$ is the single-letter index of the $\C{N}=4$ vector multiplet:
\begin{align}
    f_v=1-\frac{(1-x)(1-y)(1-z)}{(1-p)(1-q)}\ .
\end{align}
More explicitly, the matrix integral of the index \eqref{eqn: N=4 SYM (2)} can be expressed as follows: 
\begin{equation}\label{eqn: N=4 SYM (3)}
    \C{I}_{U(N)}=\frac{(\sigma,\sigma)^N (\tau,\tau)^N}{N!}\oint\prod_{i=1}^{N}du_i\frac{\prod_{i, j}\prod_{I=1}^3\Gamma(\delta_I+u_{ij},\tau,\sigma)}{\prod_{i\neq j}\Gamma(u_{ij},\tau,\sigma)}\ ,
\end{equation}
where $u_{ij}\equiv u_i-u_j$ and the chemical potentials are defined as
\begin{align}
    p=e^{2\pi i \sigma}, \quad q=e^{2\pi i \tau},\quad 
    x=e^{2\pi i \delta_1}, \quad y=e^{2\pi i \delta_2}, \quad z=e^{2\pi i \delta_3}\, \quad g_i=e^{2\pi i u_i}\ ,
\end{align}
satisfying the following constraint \cite{Choi:2018hmj, Kim:2019yrz, Amariti:2019mgp, Cassani:2021fyv}
\begin{equation}\label{eqn: N=4 constraint}
    \delta_1+\delta_2+\delta_3-\sigma-\tau=\mp1.
\end{equation}
The elliptic Gamma function $\Gamma(z,\sigma,\tau)$ and the infinite $q$-Pochhammer symbol $(z,\tau)_{\infty}$ are defined by
\begin{equation}
    \begin{split}
        \Gamma(z,\sigma,\tau)&=\prod_{m,n=0}^\infty\frac{1-e^{-2\pi iz}e^{2\pi i((m+1)\sigma+(n+1)\tau)}}{1-e^{2\pi iz}e^{2\pi i(m\sigma+n\tau)}}\ ,\\
        (z,\tau)_\infty&=\prod_{n=0}^\infty(1-e^{2\pi iz}e^{2\pi in\tau})\ .
    \end{split}
\end{equation}

Using the saddle point approximation of the matrix integral \eqref{eqn: N=4 SYM (3)}, one can find the free energy of the index in large $N$. This could be achieved by evaluating the so-called parallelogram ansatz \cite{Choi:2021rxi, Choi:2023tiq}. In particular, the eigenvalues of the matrix integral are uniformly distributed on a parallelogram region made by two edges $\sigma$ and $\tau$ (in continuum large $N$ limit)
\begin{equation}\label{eqn: Ansatz}
    u(x,y)=\sigma x+\tau y\ ,
\end{equation}
with the area density $\rho(x,y)$ satisfying $\int dx dy \rho(x,y)=1$ with $-\frac{1}{2}<x,y<\frac{1}{2}$.\footnote{The specific form of ansatz depends on the type of the gauge group. For $SO(2N)$, $SO(2N+1)$ and $Sp(N)$, the ansatz becomes $u(x,y)=\sigma x+\tau y$ with $0<x<\frac{1}{2}$ and $-\frac{1}{2}<y<\frac{1}{2}$ \cite{Choi:2023tiq}.} For a detailed derivation of how the ansatz solves the saddle point equation of the matrix integral \eqref{eqn: N=4 SYM (3)}, we refer to Appendix~\ref{sec: appendix A}.

To demonstrate that this ansatz is indeed a saddle, we reformulate the index by using the $SL(3,\mathbb{Z})$ modular properties of the elliptic gamma function \cite{Felder_2000}
\begin{align}\label{eqn: SL3Z}
    \begin{split}
        \Gamma(z,\sigma,\tau)&=e^{-\pi i P_+(z,\sigma,\tau)}\Gamma\left(-\frac{z+1}{\sigma};-\frac{1}{\sigma},-\frac{\tau}{\sigma}\right)\Gamma\left(\frac{z}{\tau};-\frac{1}{\tau},\frac{\sigma}{\tau}\right)\ ,\\
        \Gamma(z,\sigma,\tau)&=e^{-\pi i P_-(z,\sigma,\tau)}\Gamma\left(-\frac{z}{\sigma};-\frac{1}{\sigma},-\frac{\tau}{\sigma}\right)\Gamma\left(\frac{z-1}{\tau};-\frac{1}{\tau},\frac{\sigma}{\tau}\right)\ ,
    \end{split}
\end{align}
where $P_{\pm}(z,\sigma,\tau)$ is given by
\begin{align}
    P_{\pm}(z,\sigma,\tau)&=\frac{z^3}{3\sigma\tau}-\frac{\sigma+\tau\mp 1}{2\sigma\tau}z^2+\frac{\sigma^2+\tau^2+3\sigma\tau\mp3\sigma\mp3\tau+1}{6\sigma\tau}z\nonumber\\
    &\pm \frac{1}{12}(\sigma+\tau\mp1)\left(\frac{1}{\sigma}+\frac{1}{\tau}\mp 1\right)\ .
\end{align}
The two identities in \eqref{eqn: SL3Z} are related through complex conjugation, reparameterizing $(-z^*,-\sigma^*,-\tau^*)$ as $(z,\sigma,\tau)$. Note that we assumed $\Im\left(\frac{\sigma}{\tau}\right)>0$ without lose of generality; for $\Im\left(\frac{\sigma}{\tau}\right)<0$ one can use the identities with $\sigma$ and $\tau$ interchanged. 

Applying the modular properties \eqref{eqn: SL3Z} to \eqref{eqn: N=4 SYM (3)} and evaluating the ansatz, we obtain
\begin{align}\label{eqn: BH entropy}
    \log\C{I}_{U(N)}\sim-\pi i N^2\frac{ \delta_1\delta_2\delta_3}{\sigma\tau}\ .
\end{align}
which precisely matches with the entropy function of the BPS black holes in $AdS_5\times S^5$ \cite{Hosseini:2017mds}. It is important to note that the parallelogram ansatz is valid only when the following two conditions are met:
\begin{equation}\label{eqn: ansatz condition (1)}
  {\rm Im}\left(\frac{\sigma-\delta_I}{\tau}\right)<0\ ,\ \
  {\rm Im}\left(\frac{1+\delta_I}{\tau}\right)<0\ ,\ \
  {\rm Im}\left(\frac{1-\tau+\delta_I}{\sigma}\right)<0\ ,\ \
  {\rm Im}\left(\frac{\delta_I}{\sigma}\right)>0\ ,
\end{equation}
for $\delta_1+\delta_2+\delta_3-\sigma-\tau=-1$ and
\begin{equation}\label{eqn: ansatz condition (2)}
  {\rm Im}\left(\frac{\delta_I-\tau}{\sigma}\right)<0\ ,\ \
  {\rm Im}\left(\frac{1-\delta_I}{\sigma}\right)<0\ ,\ \
  {\rm Im}\left(\frac{1+\sigma-\delta_I}{\tau}\right)<0\ ,\ \
  {\rm Im}\left(\frac{\delta_I}{\tau}\right)<0\ ,
\end{equation}
and for $\delta_1+\delta_2+\delta_3-\sigma-\tau=+1$. It turns out that the parameter space of chemical potentials allowed by these inequalities is smaller than that required for CTC-free BPS black holes. This could be due to the limitation of the ansatz for the matrix integral or it could have a physical significance, which is beyond our current understanding. We also note that, throughout the paper, we will focus on the first branch with \eqref{eqn: ansatz condition (1)} for theories living on the boundary of $AdS_5$.

\subsection{Black hole entropy from the giant graviton expansion}\label{sec: Black hole entropy from the giant graviton expansion}
The GG-expansion of $\C{N}=4$ $U(N)$ SYM takes the following form \cite{Imamura:2021ytr, Gaiotto:2021xce}:
\begin{align}\label{eqn: N=4 SYM multiple-sum}
    \frac{\C{I}_{U(N)}}{\C{I}_{U(\infty)}}=\sum_{m_1,m_2,m_3=0}^{\infty}x^{m_1N}y^{m_2N}z^{m_3N} \T{\C{I}}_{(m_1,m_2,m_3)}\ ,
\end{align}
where $\C{I}_{U(\infty)}$ is the supergravity index in $AdS_5\times S^5$ \cite{Kinney:2005ej}. The giant gravitons are D3-branes wrapping three-cycles $S^3\subset S^5$. The internal manifold $S^5$ can be represented by three complex coordinates $X$, $Y$, and $Z$ in $\mathbb{C}^3$, which satisfy $|X|^2+|Y|^2+|Z|^2=1$. The different three-cycles, denoted as $S_1$, $S_2$, and $S_3$, represent intersections between $S^5$ and the $X=0$, $Y=0$, and $Z=0$ planes, respectively. Then, each term in the summation corresponds to a correction to the index resulting from D3-branes wrapping around $S_{I=1,2,3}$ with wrapping numbers $m_{I=1,2,3}$ and their open string excitations. The terms $x^{m_1N}y^{m_2N}z^{m_3N}$ originate from the classical action of the D3-brane configuration.

The quantity $\T{\C{I}}_{(m_1,m_2,m_3)}$ is defined as the index of $G=U(m_1)\times U(m_2)\times U(m_3)$ gauge theory realized on the $m_I$ coincident wrapping D3-branes. Therefore, the giant graviton index $\T{\C{I}}_{(m_1,m_2,m_3)}$ can be written as a $4d$-$2d$ coupled matrix integral
\begin{align}\label{eqn: 4d-2d index}
    \T{\C{I}}_{(m_1,m_2,m_3)}=\int_G \prod_{I=1}^{3}dg_I\;F_I^{4d}\cdot F_{I,I+1}^{2d}\ ,
\end{align}
where $\int dg_I$ is the Haar measure for $U(m_I)$ gauge group and $I+3$ is identified with $I$. The worldvolume of the wrapping D3-brane is $S^3\times\mathbb{R}$, which is equivalent to that of the boundary theory in $AdS_5$, where $\C{N}=4$ SYM with $U(N)$ gauge group lives. In other words, the functions $F_I^{4d}$ correspond to the contribution from the $\C{N}=4$ SYM with $U(m_I)$ gauge group. The functions $F^{2d}_{I,I+1}$ represent the contributions from the $2$-dimensional bifundamental hypermultiplets, which can be obtained by analyzing the open string states on the intersection of the D3-branes wrapped around two distinct three-cycles. 

We emphasize that the giant graviton lives on $S^3\subset S^5$, not on the boundary of $AdS_5$. Geometrically, this implies that the role of internal and spacetime symmetries are partly exchanged: two of the three $R$ charges play the role of angular momentum of wrapped D3-brane, while the remaining $R$ charge and two angular momenta inside $AdS_5$ serve as the `new' $R$ charges in the theory of the giant graviton. For instance, the symmetry generators acting on the $S_1$ three-cycle are related to those acting on the boundary theory via the involution map \cite{Arai:2019xmp}:
\begin{equation}\label{eqn: map}
    (J_1,J_2,Q_1,Q_2,Q_3)\;\xrightarrow{\sigma_1}\;(Q_2,Q_3,-Q_1,J_1,J_2)\ .
\end{equation}
In terms of fugacities, we write
\begin{equation}
    (p,q,x,y,z)\;\xrightarrow{\sigma_1}\;(y,z,x^{-1},p,q)\ ,
\end{equation}
which leads us to $\sigma_1 f_v=1-\frac{(1-x^{-1})(1-p)(1-q)}{(1-y)(1-z)}$. With this regard, the matrix integral \eqref{eqn: 4d-2d index} can be written more explicitly as
\begin{equation}
    \T{\C{I}}_{(m_1,m_2,m_3)}=\int_G\prod_{I=1}^3dg_I\;\text{PE}\bigg[\sum_{I=1}^3\sigma_If_v(p,q,x,y,z)\chi_{adj}^{U(m_I)}(g_I)+\sum_{I=1}^3f_{I,I+1}\chi^{(I,I+1)}\bigg]\ ,
\end{equation}
where the label $I=1,2,3$ is a cyclic variable, $I+3\sim I$ and $\chi^{(m_I,m_{I^\prime})}$ are the bifundamental characters 
\begin{equation}
    \chi^{(m_I,m_{I+1})}=\chi_{fund}^{U(m_I)}\chi_{\B{fund}}^{U(m_{I+1})}+\chi_{\B{fund}}^{U(m_I)}\chi_{fund}^{U(m_{I+1})}.
\end{equation}
The single letter index of the bifundamental contributions $f_{I,I+1}$ is given by \cite{Imamura:2021ytr}
\begin{equation}
    f_{I,I+1}=\left(\frac{x_{I+2}}{pq}\right)^{\frac{1}{2}}\frac{(1-p)(1-q)}{1-x_{I+2}}\ ,
\end{equation}
where $x_{I=1,2,3}=x,y,z$.

In practice, it is hard to perform the formal integral $\T{\mathcal{I}}_{(m_1,m_2,m_3)}$ for non-zero wrapping numbers $m_1$, $m_2$, and $m_3$ in general. Remarkably, it was suggested \cite{Gaiotto:2021xce, Imamura:2022aua,Fujiwara:2023bdc} that the multiple expansion \eqref{eqn: N=4 SYM multiple-sum} for $\C{N}=4$ SYM can be simplified to a one-parameter expansion, for instance, in $x$:
\begin{align}\label{eqn: N=4 SYM simple-sum}
    \frac{\C{I}_{U(N)}}{\C{I}_{U(\infty)}}=\sum_{m=0}^{\infty}x^{mN}\T{\C{I}}_{(m,0,0)}\ .
\end{align}
This means we can generate the full index from the contribution of a single type of giant graviton (without $2d$ bifundamental contributions), which reduces the complexity of the matrix integral \eqref{eqn: 4d-2d index}. The reduction of the multiple-sum to a simple-sum is possible because the functions $\T{\C{I}}_{(m_1,m_2,m_3)}$ are not analytic on some domain in the parameter space of fugacities. By choosing suitable expansion variables, one can eliminate some contributions of giant gravitons including bifundamental fields. This phenomenon is called ``wall-crossing'' \cite{Gaiotto:2021xce} or ``decoupling'' \cite{Fujiwara:2023bdc}.

The decoupling in $\C{N}=4$ SYM can be described as follows. The fugacities of the index are expressed as $(q, p, x, y, z)$ with the constraint $qp=xyz$. We introduce the degree $d$ for each fugacity, which can be freely chosen as long as $d\geq 0$ and satisfies the constraint $d_q+d_p=d_x+d_y+d_z$:
\begin{align}
    q=qt^{d_q}, \quad p=pt^{d_p}, \quad x=xt^{d_x}, \quad y=yt^{d_y}, \quad z=zt^{d_z}\ .
\end{align}
After redefining the fugacities, the index can be viewed as a series expansion with respect to the variable $t$.
The upshot of \cite{Fujiwara:2023bdc} is that the index $\T{\C{I}}_{(m_1,m_2,m_3)}$ becomes identically zero if a single-letter index includes infinitely many terms with negative degrees. For example, one can choose the degrees as follows
\begin{align}\label{eqn: degree}
    \textrm{deg}(p,q,x,y,z)= (1,1,0,1,1).
\end{align}
In this case, the single-letter index for $\T{\C{I}}_{(0,m_2,0)}$ is $\sigma_2 f_v=1-\frac{(1-y^{-1})(1-p)(1-q)}{(1-x)(1-z)}$, which contains infinitely many terms with negative power since $y^{-1}x^n$ is degree $-1$. Therefore, $\T{\C{I}}_{(0,m_2,0)}$ becomes identically zero and the same happens for $\T{\C{I}}_{(0,0,m_3)}$. 
On the other hand, the single-letter index for $\T{\C{I}}_{(m_1,0,0)}$, \textit{i.e.} $\sigma_1 f_v=1-\frac{(1-x^{-1})(1-p)(1-q)}{(1-y)(1-z)}$, does not have terms with negative degree. Therefore, only indices of the form $\T{\C{I}}_{(m_1,0,0)}$ survive, and the GG-expansion can be simplified to the reduced-sum as in \eqref{eqn: N=4 SYM simple-sum}. If one instead chooses degree as $\textrm{deg}(p,q,x,y,z)=(\frac{3}{2},\frac{3}{2},1,1,1)$, all the giant graviton index does not vanish and the GG-expansion is written as a multiple sum as in \eqref{eqn: N=4 SYM multiple-sum}.
Thus, the GG-expansion is not uniquely written.

Now we can reproduce the $AdS_5$ black hole entropy with general charges \eqref{eqn: BH entropy} from the simple-sum \eqref{eqn: N=4 SYM simple-sum} of GG-expansion. As discussed earlier, the worldvolume theory of the giant graviton wrapped around three-cycle $S_1$ with wrapping number $m$ is described by $\C{N}=4$ SYM with $U(m)$ gauge group, except for the relabeled chemical potential under the map $\sigma_1$, \textit{i.e.} $\T{\C{I}}_{(m,0,0)}=\sigma_1\C{I}_{U(m)}$. Therefore, the matrix integral for the reduced-sum is given by
\begin{align}\label{eqn: N=4 GGE}
    \T{\C{I}}_{(m,0,0)}\sim\oint\prod_{i=1}^{m}du_i \prod_{i\neq j}\frac{\Gamma(u_{ij}-\delta_1,\delta_2,\delta_3)\Gamma(u_{ij}+\sigma,\delta_2,\delta_3)\Gamma(u_{ij}+\tau,\delta_2,\delta_3)}{\Gamma(u_{ij},\delta_2,\delta_3)}\ .
\end{align}
Note that the constraint \eqref{eqn: N=4 constraint} now flips the sign on the right-hand side, \textit{i.e.} $\sigma+\tau+(-\delta_1)-\delta_2-\delta_3=+1$. The matrix integral \eqref{eqn: N=4 GGE} of the giant graviton index can be solved using the parallelogram ansatz, which takes the following form
\begin{equation}\label{eqn: Ansatz (2)}
    u=x\delta_2+y\delta_3, \qquad -\frac{1}{2}<x<\frac{1}{2} \quad \text{and } -\frac{1}{2}<y<\frac{1}{2}\ .
\end{equation}
with the area density $\rho(x,y)$ satisfying $\int dxdy \rho(x,y)=1$. It gives
\begin{equation}
    \log \T{\C{I}}_{(m,0,0)}\Big|_{\sigma+\tau+(-\delta_1)-\delta_2-\delta_3=+1}\sim-i\pi m^2\frac{(-\delta_1)\sigma\tau}{\delta_2\delta_3}\ .
\end{equation}
The parallelogram ansatz \eqref{eqn: Ansatz (2)} for the theory on giant graviton \eqref{eqn: N=4 GGE} is only valid if the chemical potentials satisfy the following conditions (including the assumption $\Im\big(\frac{\delta_2}{\delta_3}\big)>0$)
\begin{equation}\label{eqn: ansatz condition on GG}
  \Im\left(\frac{\delta_I^{'}-\delta_3}{\delta_2}\right)<0\ ,\ \
  \Im\left(\frac{1-\delta_I^{'}}{\delta_2}\right)<0\ ,\ \
  \Im\left(\frac{1+\delta_2-\delta_I^{'}}{\delta_3}\right)<0\ ,\ \
  \Im\left(\frac{\delta_I^{'}}{\delta_3}\right)<0\ ,
\end{equation}
where $\delta_{I}^{'}=(\sigma,\tau,-\delta_1)$. Note that we used the inequalities \eqref{eqn: ansatz condition (2)} since the relation between chemical potentials was changed into $\sigma+\tau+(-\delta_1)-\delta_2-\delta_3=+1$.

The GG-expansion in large wrapping number $m$ is now given by
\begin{align}
    \frac{\C{I}_{U(N)}}{\C{I}_{U(\infty)}}\sim \sum_{m=0}^{\infty} e^{2\pi i \delta_1 m N}e^{-i\pi m^2\frac{(-\delta_1)\sigma\tau}{\delta_2\delta_3}}\ .
\end{align}
For the next step, we assume that the summation can be approximated as the integral:
\begin{align}\label{eqn: sum to integral}
    \sum_{m=0}^{\infty} e^{2\pi i \delta_1 m N}e^{-i\pi m^2\frac{(-\delta_1)\sigma\tau}{\delta_2\delta_3}}\sim \int_0^{\infty} dm\;e^{2\pi i \delta_1 m N}e^{-i\pi m^2\frac{(-\delta_1)\sigma\tau}{\delta_2\delta_3}}\ .
\end{align}
In general, approximating the summation into the integral is not permissible for the integrand with an oscillating phase. This approximation can be justified only when there is a fine-tuned cancellation between adjacent terms in the summation. Assuming that the cancellation occurs, let us just proceed to maximize the integral above. The black hole free energy is then simply obtained by a saddle point approximation with respect to the wrapping number $m$:
\begin{equation}\label{eqn: N=4 free energy}
    \log \C{I}_{U(N)}\sim 2\pi i \delta_1 m N- \pi i m^2\frac{(-\delta_1)\sigma\tau}{\delta_2\delta_3}\Bigg|_{m=m_*}=-\frac{\pi i N^2 \delta_1\delta_2\delta_3}{\sigma\tau}\ ,
\end{equation}
where $m_*\equiv-\frac{\delta_2\delta_3 N}{\sigma \tau}$. Since the extremized wrapping number $m_*$ is in general complex, a generic black hole cannot be seen as a bound state of D3-branes; rather, it can be understood as a coherent state of D3-branes. However, as will be explicitly shown, for a small black hole, the imaginary part of $m_*$ vanishes, \textit{i.e.} the extremized wrapping number is real. Therefore, a small black hole can be described as a bound state of D-branes. This also justifies the approximation \eqref{eqn: sum to integral} without any cancellation mechanism at leading order in $N$ for the small black hole.

\subsection{Black hole with equal charges}\label{sec: Black hole with equal charges}
In this subsection, we consider the black holes with equal charges $Q_1=Q_2=Q_3=Q$ (or, equivalently, $\delta_1=\delta_2=\delta_3=\delta$), which shows several interesting features.

\paragraph{Instability and CTC}
For $\delta_1=\delta_2=\delta_3\equiv\delta$ limit, the conditions \eqref{eqn: ansatz condition on GG} simplify to\footnote{We have ignored the Cartan part in the matrix integral \eqref{eqn: N=4 GGE}, which is the order of $O(N)$ and hence subleading with respect to the free energy $O(N^2)$. However, we should be more careful to treat the Cartan part because it diverges $\propto \left(\frac{1}{(1-e^{2\pi i (-\delta_1+\delta_2)})(1-e^{2\pi i (-\delta_1+\delta_3)})}\right)^{n}$ in the equal chemical potential limit $\delta_1=\delta_2=\delta_3$ for the dual black holes with equal charges $Q_1=Q_2=Q_3$. This limit should be understood as computing the integral for different $\delta$'s and then taking them equal after integration; the poles in the Cartan part will be canceled after summing all the terms appearing in the GG-expansion \cite{Choi:2022ovw, Imamura:2021ytr}.}
\begin{equation}\label{eqn: equal delta condition of GGE}
  \Im\left(\frac{\sigma}{\delta}\right)<0,\qquad
  \Im\left(\frac{\tau}{\delta}\right)<0\ .
\end{equation}
To our surprise, if we parameterize the dual black hole solution with equal charges by two variables $a,b$, adopted in \cite{Cabo-Bizet:2018ehj, Chong:2005hr}
\begin{equation}\label{eq: on-shell chemical potential}
    \begin{split}
        \sigma&=\frac{(a-1) \left(b-i \sqrt{a b+a+b}\right)}{2 \left(a b+i (a+b+1) \sqrt{a b+a+b}+a+b\right)}, \\
        \tau&=\frac{(b-1) \left(a-i \sqrt{a b+a+b}\right)}{2 \left(a b+i (a+b+1) \sqrt{a b+a+b}+a+b\right)}, \\
        \delta&= \frac{\left(a-i \sqrt{a b+a+b}\right) \left(b-i \sqrt{a b+a+b}\right)}{2 \left(a b+i (a+b+1) \sqrt{a b+a+b}+a+b\right)}\ ,
    \end{split}
\end{equation}
the conditions \eqref{eqn: equal delta condition of GGE} for the parallelogram ansatz to be valid on the giant graviton is written as $a+b+ab>0$, which is the same as the CTC-free condition \cite{Cabo-Bizet:2018ehj, Chong:2005hr}. 

Another interesting observation is that the constraints \eqref{eqn: equal delta condition of GGE} are also related to the stability conditions of the Euclidean BPS black holes against the condensation of the D3-branes instantons \cite{Aharony:2021zkr} wrapped around a three-cycle $S^3\subset S^5$ and localized in time-direction in $AdS_5$. A gravitational background, such as a black hole, is stable only if
\begin{align}\label{eqn: D3 instability}
    \Im S_{\textrm{D3}}^{\phi}=2\pi N\Im\left(\frac{\delta}{\sigma}\right)>0,\qquad \Im S_{\textrm{D3}}^{\psi}=2\pi N\Im\left(\frac{\delta}{\tau}\right)>0\ ,
\end{align}
where the $\phi,\psi$ are the angular coordinates in $AdS_5$. In other words, the validity of the parallelogram ansatz in the GG-expansion translates into the instability of wrapping D3-branes.

\paragraph{Small black hole}
Here, we consider a small black hole limit, where $\frac{Q}{N^2}\ll 1$ (and $Q\gg J$).
For simplicity, let us also consider the equal angular momenta $J_1=J_2=J$ (or, equivalently $\sigma=\tau$). The entropy is given by the following Legendre transformation with the constraint $3\delta-2\tau=-1$,
\begin{equation}\label{eqn: Legendre transform}
    S(Q,J)=-\frac{\pi iN^2\delta^3}{\tau^2}-2\pi i\cdot 3\delta Q-2\pi i\cdot2\tau J\ .
\end{equation}

Extremizing $S(Q,J)$ with respect to $\delta$ and taking only the real part precisely yields the Bekenstein-Hawking entropy of the dual black hole. In particular, taking the small black hole limit corresponds to taking $\delta\to i0$ and $\tau\to\frac{1}{2}$. The extremized wrapping number is then given by
\begin{equation}\label{eqn: real wrapping number}
    m_*=-\frac{\delta^2}{\tau^2}N\sim\frac{2Q}{N}\in\mathbb{R}\ .
\end{equation}
Note that since $\delta$ is purely imaginary, there is no oscillating phase in \eqref{eqn: sum to integral}, which means that one can safely approximate the summation to the integral without requiring any cancellation mechanism. 
This suggests that the small black hole in AdS can be explained as a bound state of D-branes with the well-defined wrapping number \eqref{eqn: real wrapping number}. 
By studying the giant graviton index $\T{\mathcal{I}}_{(m_{*},0,0)}$, we are able to understand the small black hole physics.

Furthermore, it is suggestive that an effective $2d$ description emerges, which leads to the Cardy formula that captures the entropy of the small black hole as in \cite{Strominger:1996sh}.\footnote{The authors of \cite{Strominger:1996sh} count flat black hole microstates by considering the effective 2d description of D1-D5 system. A small $AdS_5$ black hole can be regarded as a flat black hole where AdS space acts as the IR regulator as well as a bound state of D-branes.} 
The relevant idea was first given by \cite{Kinney:2005ej} and the GG-expansion provides a concrete manifestation of it. The single-letter index $\sigma_1 f_v$ of the giant graviton index with equal charges, \textit{i.e. } $x=y=z$, is given by 
\begin{equation}
    \sigma_1f_v=1+\frac{1}{x}\frac{(1-p)(1-q)}{(1-x)}=1+\sqrt{\frac{x}{pq}}\frac{(1-p)(1-q)}{(1-x)}\ .
\end{equation}
Indeed, this expression is identical to the single-letter index of $2d$ $\C{N}=(4,4)$ vector multiplet living on a circle.

Let us now explicitly compute the entropy and central charge of the effective $2d$ system. Firstly, by plugging the wrapping number \eqref{eqn: real wrapping number} into \eqref{eqn: Legendre transform} and extremizing it, the entropy of $2d$ system emerging from the wrapped D3-branes is given by
\begin{equation}\label{eqn: small black hole entropy}
    S(Q)\approx \frac{2\pi\sqrt{2}Q^{3/2}}{N}\ ,
\end{equation}
where $\delta=i\sqrt{\frac{Q}{2N^2}}$. One can obtain the central charge $c$ as follows. For a thermodynamic system, the Cardy formula can be written as
\begin{align}\label{eq: cardy with temp}
    S=\frac{\pi^2}{3}cT\ .
\end{align}
In the case of BPS black holes, of course, temperature vanishes. However, one can introduce the so-called Frolov-Thorne temperature $T_{FT}$ \cite{Frolov:1989jh,Cabo-Bizet:2018ehj} which is a renormalized temperature defined as
\begin{align}\label{eqn: Frolov Thorne temp}
    T_{FT}=\lim_{T\rightarrow 0}\frac{T}{\Phi-1}=-\frac{1}{2\pi i \delta}\ .
\end{align}
where $\Phi$ is the chemical potential of the electric charge. By combining \eqref{eqn: small black hole entropy} $\sim$ \eqref{eqn: Frolov Thorne temp}, the central charge $c$ of the system is given by
\begin{align}
    c=12\frac{Q^2}{N^2}\ .
\end{align}
This computation is consistent with the expectation given by \cite{Kinney:2005ej}.

\section{Other examples}\label{sec: Other examples}
In this section, we study the GG-expansion to analyze several well-known holographic models for large $N$.

\subsection{Half-BPS $2d$ surface defect in $\C{N}=4$ SYM}
In this subsection, we consider the index of the $\C{N}=4$ $U(N)$ SYM with a half-BPS $2d$ surface defect insertion studied in \cite{Nakayama:2011pa}. This theory can be described by a $4d$-$2d$ coupled system, in which the bulk $\C{N}=4$ SYM coupled to a $2d$ $(4,4)$ hypermultiplet. The defect, which is a co-dimension two object extended along the time and one spatial direction, breaks some amount of supersymmetry. It preserves the following superconformal algebra:
\begin{align}
    u(1)_A \ltimes (\mathfrak{psu}(1,1|2)\times \mathfrak{psu}(1,1|2)) \ltimes u(1)_C \subset \mathfrak{psu}(2,2|4)\ .
\end{align}
In particular, the bosonic part is $\mathfrak{sl}(2,\mathbb{R})\times \mathfrak{sl}(2,\mathbb{R})\times \mathfrak{u}(1)\times \mathfrak{su}(2)\times \mathfrak{su}(2)\times \mathfrak{u}(1)$. Since the number of Cartans still remains unchanged, one can keep all the fugacities and define the superconformal index in the same way as \eqref{eqn: N=4 SYM (1)}.\footnote{The half-BPS surface defect differs from the Gukov-Witten defect \cite{Gukov:2006jk}. The $4d$-$2d$ description of the GW defect includes a $2d$ $(4,4)$ vector multiplet in addition to the hypermultiplet. However, it gives only $O(1)$ contributions and is therefore negligible in large $N$ analysis \cite{Chen:2023lzq}.} 

In AdS/CFT, the boundary theory with the half-BPS surface defect can be constructed by $N$ coincident D3-branes intersecting with a probing D3-brane \cite{Constable:2002xt}. More precisely, %
In the near-horizon limit, the geometry of the defect becomes $AdS_3\times S^1$ inside $AdS_5\times S^5$, where $S^1$ is one of the great circles in $S^5$. 

The index is written as 
\begin{equation}
    \begin{split}
        &\C{I}^{\text{defect}}_{U(N)}\\
        &\;\;= \int_{U(N)}dg\;\textrm{PE}\left[f_v(p,q,x,y,z)\chi_{adj}^{U(N)}(g)\right]\textrm{PE}\left[\sqrt{\frac{q}{xy}}\frac{(1-x)(1-y)}{(1-q)}(\Tr U +\Tr U^{\dagger})\right]\\
        &\;\;\sim\oint\prod_{i=1}^{N}du_i\prod_{i\neq j}\frac{\prod_{I=1}^3\Gamma(\delta_I+u_{ij},\tau,\sigma)}{\Gamma(u_{ij},\tau,\sigma)}
        \frac{\theta(u_i+\frac{\delta_1-\delta_2+\tau-1}{2},\tau) \theta(-u_i+\frac{\delta_1-\delta_2+\tau-1}{2},\tau)}{\theta(u_i+\frac{\delta_1+\delta_2+\tau-1}{2},\tau) \theta(-u_i+\frac{\delta_1+\delta_2+\tau-1}{2},\tau)}\ ,
    \end{split}
\end{equation}
where $\theta(z,\tau)= (z,\tau)_{\infty}(\tau-z,\tau)_{\infty}$ is a $q$-theta function. Similar to the elliptic Gamma functions, the $q$-theta function $\theta(z,\tau)$ also exhibits an $SL(2,\mathbb{Z})$ transformation property:
\begin{align}\label{eqn: SL(2,Z)}
    \theta(z,\tau)=e^{-\pi i B(z,\tau)}\theta\Big(\frac{z}{\tau},-\frac{1}{\tau}\Big)\ ,
\end{align}
where the prefactor $B(z,\tau)$ is given by
\begin{align}
    B(z,\tau)\equiv \frac{z^2}{\tau}+z\left(\frac{1}{\tau}-1\right)+\frac{1}{6}\left(\tau+\frac{1}{\tau}\right)-\frac{1}{2}\ .
\end{align}
Using the $SL(3,\mathbb{Z})$ and $SL(2,\mathbb{Z})$ modular properties, respectively, one can show that the parallelogram ansatz \eqref{eqn: Ansatz} satisfies the saddle point equation. On the other hand, collecting all the prefactors, the free energy is given by
\begin{align}\label{eqn: defect freeE}
    \log \C{I}^{\text{defect}}_{U(N)} = -\frac{\pi i N^2 \delta_1\delta_2\delta_3}{\sigma\tau}+\frac{2\pi i N \delta_1\delta_2}{\tau}\ .
\end{align}
The result is the same as that obtained in \cite{Chen:2023lzq}, which shows that the defect index can be used as an order parameter for the deconfinement phase transition. Note that we have omitted the Cartan part contributions in computing the free energy even though it is of order $N$ since it does not affect the expectation value of the defect, defined as $\frac{\log \C{I}^{\text{defect}}_{U(N)} }{\log \C{I}_{U(N)} }$.

Now we consider the reduced GG-expansion of the half-BPS defect index. We suggest the GG-expansion as\footnote{In \cite{Gaiotto:2021xce}, the authors also studied the GG-expansion of this system. They explicitly presented $\T{\C{I}}^{\text{defect}}_{(1,0,0)}$ only, from which one can also infer the GG-expansion formula.}
\begin{align}
    \frac{\C{I}^{\text{defect}}_{U(N)}}{\C{I}^{\text{defect}}_{U(\infty)}}=\sum_{m=0}^{\infty}x^{mN} \T{\C{I}}^{\text{defect}}_{(m,0,0)}\ ,
\end{align}
The giant graviton index $\T{\C{I}}^{\text{defect}}_{(m,0,0)}$ can be described by a $4d$-$2d$ system, which consists of $4d$ $U(m)$ theory living on $X=0$ (\textit{i.e.} $S^3\times \mathbb{R}$) and $2d$ bifundamental hypermultiplet on the intersection between the giant graviton and the probing D3-brane (\textit{i.e.} $AdS_3\times S^1\subset AdS_5\times S^5$). Note also that, while the giant graviton resides in the bulk of $AdS_5$, the surface defect D3-brane extends to the asymptotic boundary \cite{Constable:2002xt}. The giant graviton index is written as
\begin{align}
    &\T{\C{I}}^{\text{defect}}_{(m,0,0)}\nonumber\\
    &\;\;=\int dg\;\text{PE}\bigg[\sigma_1f_v(p,q,x,y,z)\chi_{adj}^{U(m)}(g)\bigg]\textrm{PE}\bigg[\sqrt{\frac{z}{x^{-1}p}}\frac{(1-x^{-1})(1-p)}{(1-z)}(\Tr U+\Tr U^\dagger)\bigg]\nonumber\\
    &\;\;\sim\oint\prod_{i=1}^{m}du_i \prod_{i\neq j}\frac{\Gamma(u_{ij}-\delta_1,\delta_2,\delta_3)\Gamma(u_{ij}+\sigma,\delta_2,\delta_3)\Gamma(u_{ij}+\tau,\delta_2,\delta_3)}{\Gamma(u_{ij},\delta_2,\delta_3)}\\
    &\hspace{2.4cm}\times\frac{\theta(u_i+\frac{\delta_1+\delta_3+\sigma-1}{2},\delta_3)\theta(-u_i+\frac{\delta_1+\delta_3+\sigma-1}{2},\delta_3)}{\theta(u_i+\frac{\delta_3-\delta_1+\sigma-1}{2},\delta_3)\theta(-u_i+\frac{\delta_3-\delta_1+\sigma-1}{2},\delta_3)}\ ,\nonumber
\end{align}
where $f_{b}=\sqrt{\frac{z}{x^{-1}p}}\frac{(1-x^{-1})(1-p)}{(1-z)}$ is the single-letter index a bifundamental hypermultiplet living at the intersection. Once again, we extract the free energy from the prefactors:
\begin{equation}
    \log \C{I}_{U(N)}^{\text{defect}}\sim 2\pi i \delta_1 m N- \pi i m^2\frac{(-\delta_1)\sigma\tau}{\delta_2\delta_3}+2\pi i m\frac{(-\delta_1)\sigma}{\delta_3}\ .
\end{equation}
By extremizing this with respect to wrapping number $m$, we obtain
\begin{align}
    \log \C{I}_{U(N)}^{\text{defect}} \sim -\pi i N^2\frac{\delta_1 \delta_2\delta_3}{\sigma\tau}+2\pi i N \frac{\delta_1\delta_2}{\tau}+O(1)\ ,
\end{align}
which is exactly the same as \eqref{eqn: defect freeE}.

\subsection{Orbifold quiver gauge theories}
Consider a theory \cite{Kachru:1998ys, Lawrence:1998ja, Nakayama:2005mf} that is constructed on the worldvolume of $N(\equiv k\B{N})$ D3-branes at the origin of the orbifold $\mathbb{C}^3/\mathbb{Z}_k$ background.
Since the orbifold action commutes with taking the near-horizon limit, the dual holographic geometry is of the form $AdS_5\times S^5/\mathbb{Z}_k$.
We focus on the orbifold action $\mathbb{Z}_k\subset SU(3)\subset SU(4)_R$\footnote{One can also analyze the case for $\mathbb{Z}_k\subset SU(2)\subset SU(4)_R$, which preserves $\C{N}=2$ supersymmetry.} on $\mathbb{C}^3=(X,Y,Z)$, generated by
\begin{equation}\label{eqn: N=1 orbifold action (1)}
    \begin{pmatrix}
        e^{\frac{2\pi i}{k} \alpha_1}&0&0\\
        0&e^{\frac{2\pi i}{k} \alpha_2}&0\\
        0&0&e^{\frac{2\pi i}{k} \alpha_3}
    \end{pmatrix}\ ,
\end{equation}
with $\alpha_1+\alpha_2+\alpha_3=0$ (mod $k$) with non-zero $\alpha_{1,2,3}$. By construction, the boundary theory is obtained from the $\C{N}=4$ $U(N)$ SYM by the $\mathbb{Z}_k$ orbifold projection with the generator 
\begin{equation}\label{eqn: N=1 orbifold action (2)} 
    \exp\left(\frac{2\pi i}{k}(\alpha_1 Q_1+\alpha_2 Q_2+\alpha_3 Q_3)\right)\ .
\end{equation}
The resulting theory is the $\C{N}=1$ quiver gauge theory with the $U(\B{N})^k$ gauge group and various bifundamental fields. In fact, the diagonal subgroups $U(1)\subset U(\B{N})$ are IR-free and become global baryonic symmetries. However, we will focus on the sector with vanishing baryonic charges for simplicity, which is equivalent to considering the gauge group $U(\B{N})$ instead of $SU(\B{N})$ when computing the index, as we did so far.

The insertion of the generator \eqref{eqn: N=1 orbifold action (2)} (which obviously commutes with the supercharge $\C{Q}$ used to define the index) in the trace formula \eqref{eqn: N=4 SYM (1)} translates into the following $\mathbb{Z}_k$ action on the fugacities:
\begin{equation}
    (p,q,x,y,z)\;\to\;(p,q,\omega_k^{\alpha_1} x,\omega_k^{\alpha_2} y,\omega_k^{\alpha_3} z),\qquad \omega_k=e^{\frac{2\pi i}{k}}\ .
\end{equation}
The index is then given by
\begin{equation}
    \C{I}_{U(\B{N})^k}=\int_{U(\B{N})^k}dg\;\textrm{PE}\left[P_k\left(f_v(p,q,x,y,z)\chi_{adj}^{U(N)}(g)\right)\right]\ .
\end{equation}
Here, $P_k$ denotes the $\mathbb{Z}_k$ projection
\begin{equation}\label{eqn: projection}
    \begin{split}
        P_k&\left(f_v(p,q,x,y,z)\Tr U\Tr U^{\dagger }\right)\\
        &=\frac{1}{k}\sum_{l=0}^{k-1}\left[1-\frac{(1-\omega^{\alpha_1 l}x)(1-\omega^{\alpha_2 l}y)(1-\omega^{\alpha_3 l}z)}{(1-p)(1-q)}\right]\sum_{a,b=1}^k\Tr U_a\Tr U_b^\dagger\omega^{(a-b)l}\ .
    \end{split}
\end{equation}
where $a$ labels the different gauge nodes. In terms of elliptic gamma functions, the index is given by
\begin{equation}
    \C{I}_{U(\B{N})^k}\sim\oint \prod_{a=1}^k\prod_{i=1}^{\B{N}}du^{(a)}_i \frac{\prod_{i,j=1}^{\B{N}}\Gamma(u_{ij}^{(a,a+\alpha_1)}+\delta_{1},\sigma,\tau)\Gamma(u_{ij}^{(a,a+\alpha_2)}+\delta_{2},\sigma,\tau)\Gamma(u_{ij}^{(a,a+\alpha_3)}+\delta_{3},\sigma,\tau)}{\prod_{i\neq j}^{\B{N}}\Gamma(u_{ij}^{(aa)},\sigma,\tau)}\ ,
\end{equation}
where $u_{ij}^{(a,b)}\equiv u_i^{(a)}-u_j^{(b)}$. Evaluating the parallelogram ansatz \eqref{eqn: Ansatz} under the conditions \eqref{eqn: ansatz condition (1)} yields
\begin{equation}\label{eqn: orbifold free energy}
    \log\mathcal{I}_{U(\B{N})^k}\Big|_{\delta_1+\delta_2+\delta_3-\sigma-\tau=-1}\sim
    -i\pi k \B{N}^2 \frac{\delta_1\delta_2\delta_3}{\sigma \tau} =-i\pi \frac{N^2}{k}  \frac{\delta_1\delta_2\delta_3}{\sigma \tau}\ .
\end{equation}
This is consistent with the fact that $\textrm{Vol}(S^5/\mathbb{Z}_k)=\frac{1}{k}\textrm{Vol}(S^5)$.

We now want to reproduce the result \eqref{eqn: orbifold free energy} using the GG-expansion. Similar to the previous section, the GG-expansion for the $\mathbb{C}^3/\mathbb{Z}_k$ orbifold theory can be expressed as a simple-sum if we choose the degrees \eqref{eqn: degree}. In addition, the worldvolume of D3-branes wrapped around the three-cycle $S_1$ with wrapping number $m(\equiv k\B{m})$ is also given by $S^3/\mathbb{Z}_k\times\mathbb{R}$. Therefore the theory of giant graviton is the same as $\C{N}=1$, $\mathbb{Z}_k$ orbifold theory with $U(\B{m})^k$ gauge group (up to $\sigma_1$ mapping). Hence, the GG-expansion is
\begin{equation}
    \frac{\C{I}_{U(\B{N})^k}}{\C{I}_{U(\infty)^k}}=\sum_{\B{m}=0}^\infty x^{k\B{m} \B{N}}\T{\C{I}}_{U(\B{m})^k}\ .
\end{equation}
Using the definition of $\sigma_1$ and \eqref{eqn: projection}, one can derive the following result (see Appendix~\ref{sec: appendix B})
\begin{equation}\label{eqn: Orbifold GGE}
    \begin{split}
        \T{\C{I}}_{U(\B{m})^k}&\sim\oint\prod_{a=1}^k\prod_{i=1}^{\B{m}}du_i^{(a)}\prod_{b=0}^{k-1}\prod_{c=0}^{k-1}\prod_{i,j}^{\B{m}}\frac{\Gamma(u_{ij}^{(a, [\![ a+b\alpha_2+c\alpha_3 ]\!])}+b\delta_2+c\delta_3+\sigma, k\delta_2,k\delta_3)}{\Gamma(u_{ij}^{(a, [\![ a+b\alpha_2+c\alpha_3 ]\!])}+b\delta_2+c\delta_3, k\delta_2,k\delta_3)}\\
        &\hspace{3cm}\times \frac{\Gamma(u_{ij}^{(a, [\![ a+b\alpha_2+c\alpha_3 ]\!])}+b\delta_2+c\delta_3+\tau, k\delta_2,k\delta_3)}{\Gamma(u_{ij}^{(a, [\![ a+b\alpha_2+c\alpha_3 ]\!])}+b\delta_2+c\delta_3+\tau+\sigma, k\delta_2,k\delta_3)}\ ,
    \end{split}
\end{equation}
where for an integer $l$ we define $ [\![ l ]\!]$ to be modulo $k$. The structure of the matrix integral shows that the quiver gauge theory on giant graviton consists of $k$ vector multiplets and several chiral multiplet $\Phi_\chi$, which can be in either bifundamental or adjoint representation. For instance, when $b=c=0$, the elliptic gamma function $1/\Gamma(u_{ij}^{(a,a)},k\delta_2,k\delta_3)$ corresponds to the contributions of a vector multiplet for $a$-th gauge node. Then, other elliptic gamma functions, dressed by a linear combination of various chemical potentials $\delta_2$, $\delta_3$, $\sigma$, and $\tau$ correspond to the contributions from $\Phi_\chi$.\footnote{The elliptic gamma function in denominator in the second line can be converted into numerator using the identity $\Gamma\big(u+b\delta_2+c\delta_3+\tau+\sigma,k\delta_2,k\delta_3\big)\Gamma\big(k\delta_2+k\delta_3-(u+b\delta_2+c\delta_3+\tau+\sigma),k\delta_2,k\delta_3\big)=1$}

Now, evaluating the ansatz \eqref{eqn: Ansatz (2)} gives
\begin{equation}
    \log \T{\C{I}}_{U(\B{m})^k}\Big|_{\sigma+\tau+(-\delta_1)-\delta_2-\delta_3=+1}\sim-i\pi k\B{m}^2\frac{(-\delta_1)\sigma\tau}{\delta_2\delta_3}\ ,
\end{equation}
so that
\begin{equation}
    \frac{\C{I}_{U(\B{N})^k}}{\C{I}_{U(\infty)^k}}\sim\sum_{\B{m}=0}^\infty e^{2\pi ik\delta_1\B{m}\B{N}}e^{-i\pi k \B{m}^2\frac{(-\delta_1)\sigma\tau}{\delta_2\delta_3}}\ .
\end{equation}
By approximating the summation as integral and extremizing the wrapping number, we obtain
\begin{equation}
    \log\C{I}_{U(\B{N})^k}\sim 2\pi i k \delta_1\B{m}\B{N}-i\pi k\B{m}^2\frac{(-\delta_1)\sigma\tau}{\delta_2\delta_3}\Bigg|_{\B{m}=\B{m}_*}=-i\pi k\B{N}^2\frac{\delta_1\delta_2\delta_3}{\sigma\tau}\ ,
\end{equation}
where $\B{m}_*=-\frac{\delta_2\delta_3\B{N}}{\sigma\tau}$. Therefore, we again reproduce the correct answer.

\subsection{Toric quiver gauge theories}
Consider the $\C{N}=1$ toric quiver gauge theories that can be constructed on the worldvolume of $N$ D3-branes probing a toric Calabi-Yau cone over Sasaki-Einstein $5$-manifold $C(SE_5)$, whose dual geometry is of the form $AdS_5\times SE_5$. The well-known examples of the $SE_5$ geometry includes $Y^{p,q}$ and $T^{1,1}$ \cite{Martelli:2004wu, Benvenuti:2004dy,Klebanov:1998hh}. 

\begin{figure}
    \centering
    \includegraphics[scale=0.35]{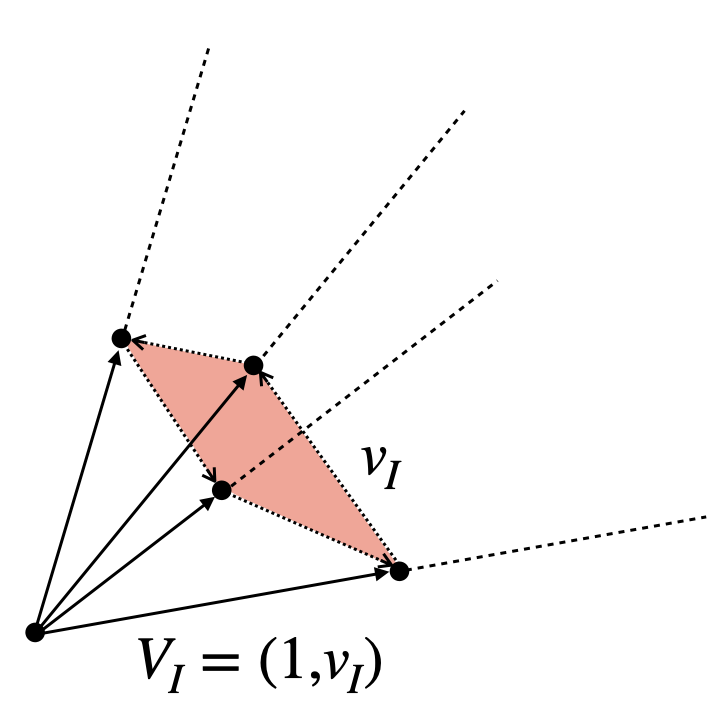}
    \caption{Toric fan}
    \label{fig: Fan and Toric diagram}
\end{figure}

The toric quiver gauge theories can be specified by toric data encoded in the toric diagram associated with the $C(SE_5)$. More precisely, the singularity of $C(SE_5)$ can be described by a fan, which is generated by $d$ primitive vectors $V_I$ in $\mathbb{R}^3$. Since the cone is Calabi-Yau threefold, these vectors can be restricted to a plane by setting the last coordinate equal to $1$, \textit{i.e.} $V_I=V_I(v_I,1)\equiv(x_I,y_I,1)$. This plane, made by integer vertices and vectors $v_I$ along the perimeter, defines a regular convex polygon called the toric diagram, as shown in Figure~\ref{fig: Fan and Toric diagram}.

There is a systematic procedure to extract the boundary theory data, including gauge group, matter contents, and superpotential interactions, from the toric data of the $C(SE_5)$ \cite{Hanany:2005ve, Franco:2005sm, Franco:2005rj, Feng:2005gw}. For example, the perimeter $d$ of the toric diagram is related to the number of anomaly-free $U(1)$ global symmetries of the boundary theory. In particular, we always have $U(1)^3$ global symmetries from the isometries of $SE_5$, which correspond to two mesonic global symmetries and one $R$-symmetry. We also find $d-3$ baryonic symmetries,\footnote{In gravity side, the gauge fields that correspond to the $d-3$ baryonic symmetries are obtained by dimensional reduction of RR four-form field in type IIB string theory on the topological three-cycles of $SE_5$.} which correspond to the number of independent topological three-cycles of $SE_5$. These three-cycles are topologically Lens spaces $S^3/\mathbb{Z}_{k_I}$ and classified by the fact that $H_3(SE_5,\mathbb{Z})\cong \mathbb{Z}^{d-3}$. Thus, there are a total of $d$ non-trivial three-cycles in $SE_5$ that are available for D3-branes to wrap around.

Now, the index of a generic $\C{N}=1$ toric quiver gauge theory is defined as
\begin{align}\label{eqn: Toric index}
    \C{I}=\Tr\bigg[(-1)^F p^{J_1}q^{J_2} \prod_{I=1}^{d}v_I^{R_I}\bigg]\ ,
\end{align}
with a constraint $pq=\prod_{I=1}^{d}v_I$ and $v_I=e^{2\pi i\delta_I}$. Here, we have defined a new basis of chemical potentials and charges by mixing the $d-3$ $U(1)$ baryonic symmetries $B_{a=1,\cdots,d-3}$, two mesonic symmetries $F_{1,2}$ and $U(1)_R$ symmetry $R_0$, such that all $R_I$ acts on the supercharge $\C{Q}$ in the same way, with normalization $[R_I,\C{Q}]=+\frac{1}{2}\C{Q}$. This particular choice of basis is natural in that the $d$ $U(1)$ symmetries with charge $R_I$ are identified with the $d$ boundary points $V_I$ of the toric diagram. In other words, we can assign a basis of global symmetries of the quiver gauge theory directly from the toric diagram. Note also that the true superconformal $R$-charge can be determined by the $a$-maximization procedure \cite{Intriligator:2003jj}, which is equivalent to the volume minimization \cite{Martelli:2005tp} via AdS/CFT.

In general, the GG-expansion is given by \cite{Fujiwara:2023bdc, Arai:2019aou}:
\begin{align}\label{eqn: d GG-expansion}
    \frac{\C{I}_N}{\C{I}_{\infty}}=\sum_{m_1,\cdots,m_d=0}^\infty v_1^{m_1}\cdots v_d^{m_d}\;\T{\C{I}}_{(m_1,\cdots,m_d)}\ ,
\end{align}
where, again, the factor $v_I^{m_I}$ comes from the classical contribution of D3-branes wrapped on a non-trivial three-cycles $S_I$ corresponding to the corner of the toric diagram. The giant graviton index $\T{\C{I}}_{(m_1,\cdots,m_d)}$ is defined as the that of $G=U(m_1)\times\cdots\times U(m_d)$ coupled to $2d$ bifundamental hypermultiplets attached at the intersections
\begin{align}
    \T{\C{I}}_{(m_1,\cdots,m_d)}&=\int_G\prod_{I=1}^d dg_I\;F_I^{4d}\cdot F_{I,I+1}^{2d}\\
    &=\int_G\prod_{I=1}^d dg_I\; \textrm{PE}\bigg[\sum_{I=1}^d P_{k_I}\left( \sigma_I f_I(p,q,v_I)\chi_{adj}^{U(m_I)}(g_I)\right)+\sum_{I,I+1}f_{I,I+1}\chi^{(m_I,m_{I+1})}\bigg]\ .\nonumber
\end{align}
where $\chi^{(m_I,m_{I^\prime})}$ are the bifundamental characters 
\begin{equation}
    \chi^{(m_I,m_{I+1})}=\chi_{fund}^{U(m_I)}\chi_{\B{fund}}^{U(m_{I+1})}+\chi_{\B{fund}}^{U(m_I)}\chi_{fund}^{U(m_{I+1})}\ .
\end{equation}
The projection $P_{k_I}$ must be needed because the non-trivial three-cycles in $SE_5$ are topologically Lens space $S^3/\mathbb{Z}_{k_I}$. 
In particular, the order of projection $k_I$ for each three-cycle $S_I$ and the single-letter index can be determined from the toric diagram:
\begin{equation}
    \sigma_If_I=1-\frac{\big(1-\frac{(w_r w_{r^\prime})^{\frac{1}{k_I}}}{pq}\big)(1-q)(1-p)}{\big(1-w_r^{\frac{1}{k_I}}\big)\big(1-w_{r^\prime}^{\frac{1}{k_I}}\big)}\ ,
\end{equation}
where we introduced the fugacities, defined by
\begin{equation}
    \omega_r=\prod_{I=1}^d v_I^{V_I\cdot g_r},\qquad \omega_{r^\prime}=\prod_{I=1}^d v_I^{V_I\cdot g_{r^\prime}}\ .
\end{equation}
The cyclic variables $r\in I+\frac{1}{2}$ and $r^\prime\in I-\frac{1}{2}$ are used to denote two adjacent edges that meet at the corner $I$ in the toric diagram. The vectors $g_r$ (and also $g_{r^\prime}$) are primitive integer vectors orthogonal to $V_{r-1/2}$ and $V_{r+1/2}$, with $g_r \cdot V_I \geq 0$ for all $I$. Similarly, the single letter index $f_{I,I+1}$ from $2d$ contributions on the intersecting three-cycles $S_I$ and $S_{I+1}$ are given by
\begin{equation}
    f_{I,I+1}=\left(\frac{w_{r}}{pq}\right)^{\frac{1}{2}}\frac{(1-q)(1-p)}{1-w_{r}}\ .
\end{equation}

The claim is that the $d$ multiple-sum in \eqref{eqn: d GG-expansion} can be simplified, at most, to a $d-2$-ple sum \cite{Fujiwara:2023bdc}: it is impossible to remove $d-3$ topological cycles, which have no geometric action, while the remaining three non-topological cycles can be reduced to a single three-cycle. For instance, if the toric diagram for $C(SE_5)$ is a triangle, then the simple-sum expansion works, similar to $\C{N}=4$ SYM. In the following, we will see how this reduced-sum of GG-expansion works and reproduces the entropy of the dual black holes.

\subsubsection{$T^{1,1}$}

\begin{figure}[t]
    \centering
    \begin{subfigure}[b]{0.48\linewidth}  
        \centering
        \includegraphics[width=0.6\linewidth]{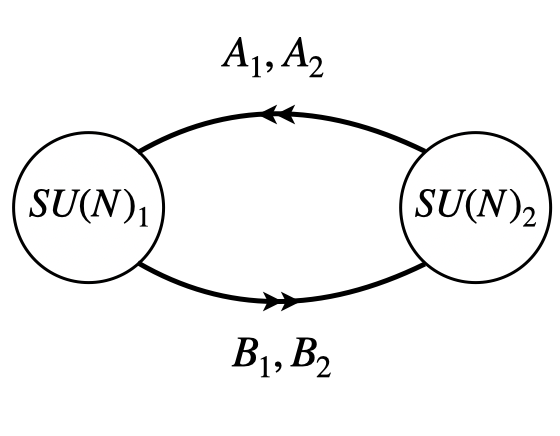}
        \caption{}
        \label{fig:A}
    \end{subfigure}
    \begin{subfigure}[b]{0.48\linewidth}
        \centering
        \includegraphics[width=0.3\linewidth]{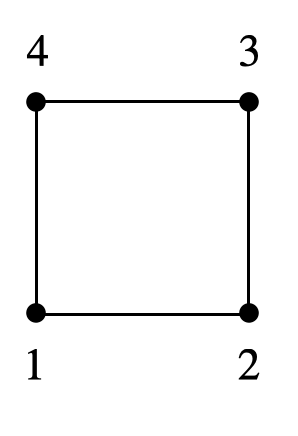}
        \caption{}
        \label{fig:B}
    \end{subfigure}
    \caption{(a) The quiver diagram of the Klebanov-Witten theory (b) The toric diagram of the Klebanov-Witten theory.}
    \label{fig: KW Quiver and Toric diagram}
\end{figure}

\begin{table}[t]
\small
    \centering
    \begin{tabular}{|c|c|c|c|c||c|c|c|c|}
    \hline
     & $U(1)_{F_1}$ & $U(1)_{F_2}$ & $U(1)_B$ & $U(1)_{R_0}$ & $U(1)_{1}$ & $U(1)_{2}$ & $U(1)_{3}$ & $U(1)_{4}$\\ \hline
     $A_1$ & $1$ & $0$ & $1$ & $1/2$ & $1$ & $0$ & $0$ & $0$\\\hline 
     $A_2$ & $-1$ & $0$ & $1$ & $1/2$ & $0$ & $1$ & $0$ & $0$\\ \hline
     $B_1$ & $0$ & $1$ & $-1$ & $1/2$ & $0$ & $0$ & $1$ & $0$\\ \hline
     $B_2$ & $0$ & $-1$ & $-1$ & $1/2$ & $0$ & $0$ & $0$ & $1$\\ \hline
    \end{tabular}
    \caption{Charge assignment of each chiral multiplet in the Klebanov-Witten theory.}
    \label{tab: Charge assignment global symmetries for KW theory}
\end{table}

Let us consider the Klebanov-Witten theory \cite{Klebanov:1998hh} whose dual geometry is $AdS_5\times T^{1,1}$. In this case, the toric Calabi-Yau cone is the conifold at which $N$ D3-branes are probing. This theory has $U(N)\times U(N)$ gauge group 
with two pairs of bifundamental chiral multiplets $A_{1,2}$ and $B_{1,2}$ transforming in the $(\bm{N},\B{\bm{N}})$ and $(\B{\bm{N}},\bm{N})$ representations. The corresponding toric diagram and quiver diagram are shown in Figure~\ref{fig: KW Quiver and Toric diagram}. Table~\ref{tab: Charge assignment global symmetries for KW theory} shows the charge assignment for each multiplet. In terms of chemical potentials, the convenient basis $R_{I=1,2,3,4}$ for the $U(1)_{F_1}\times U(1)_{F_2}\times U(1)_B\times U(1)_{R}$ global symmetries are
\begin{equation}
    \begin{split}
        \delta_1&=\xi_{F_1}+\xi_{B}+\frac{1}{4}(\tau+\sigma)\ , \qquad\delta_2=-\xi_{F_1}+\xi_{B}+\frac{1}{4}(\tau+\sigma)\ ,\\
        \delta_3&=\xi_{F_2}-\xi_{B}+\frac{1}{4}(\tau+\sigma)\ , \qquad\delta_4=-\xi_{F_2}-\xi_{B}+\frac{1}{4}(\tau+\sigma)\ ,
    \end{split}
\end{equation}
Then, the matrix integral of the index is given by
\begin{align}\label{eqn: KW index}
    \begin{split}
        &\mathcal{I}_{T^{1,1}[N]}\sim\oint\prod_{i=1}^{N}du_i^{(1)}du_i^{(2)}\frac{1}{\prod_{i\neq j}^N\Gamma\Big(u_{ij}^{(11)},\sigma,\tau\Big)\Gamma\Big(u_{ij}^{(22)},\sigma,\tau\Big)}\\
        &\;\times\prod_{i,j=1}^N\Gamma\Big(u_{ij}^{(12)}+\delta_{1},\sigma,\tau\Big)\Gamma\Big(u_{ij}^{(12)}+\delta_{2},\sigma,\tau\Big)\Gamma\Big(u_{ij}^{(21)}+\delta_{3},\sigma,\tau\Big)\Gamma\Big(u_{ij}^{(21)}+\delta_{4},\sigma,\tau\Big)\ ,
    \end{split}
\end{align}
with the constraint $\delta_1+\delta_2+\delta_3+\delta_4-\sigma-\tau=-1$. The large $N$ analysis of the index for a generic $\C{N}=1$ toric quiver gauge theories has been extensively done in \cite{Benini:2020gjh,GonzalezLezcano:2019nca,Cabo-Bizet:2020nkr}. 
In particular, in \cite{Choi:2023tiq}, the authors apply the parallelogram ansatz to solve this matrix integral in large $N$, which yields
\begin{equation}
    \log\mathcal{I}_{T^{1,1}[N]}\sim -\pi i N^2\frac{\delta_1\delta_2\delta_3+\delta_2\delta_3\delta_4+\delta_1\delta_2\delta_4+ \delta_1\delta_3\delta_4}{\sigma \tau}\ .
\end{equation}

If we choose the degrees as $\text{deg}(v_1,v_2,v_3,v_4) = \left(0,0,\frac{1}{2},\frac{1}{2}\right)$, the GG-expansion can be reduced into double-sum:
\begin{align}
    \frac{\C{I}_{T^{1,1}[N]}}{\C{I}_{T^{1,1}[N]}}=\sum_{m_1,m_2=0}^\infty v_1^{m_1N}v_2^{m_2N} \T{\C{I}}_{(m_1,m_2,0,0)}\ ,
\end{align}
where
\begin{align}
    \T{\C{I}}_{(m_1,m_2,0,0)}=\int_G dg \;\text{PE}\left( f_1\chi_{adj}^{U(m_1)}+f_2\chi_{adj}^{U(m_2)}+f_{12}\chi^{(m_1,m_2)}\right)\ ,
\end{align}
and from the toric data, we obtain
\begin{equation}
    \begin{split}
        f_1&=1-\frac{\left(1-\frac{v_3}{v_1}\right)(1-p)(1-q)}{\left(1-v_2v_3\right)\left(1-v_4v_3\right)},\quad f_2=1-\frac{\left(1-\frac{v_4}{v_2}\right)(1-p)(1-q)}{\left(1-v_3v_4\right)\left(1-v_1v_3\right)}\ ,\\
        f_{12}&=\left(\frac{v_3v_4}{pq}\right)^{\frac{1}{2}}\frac{(1-p)(1-q)}{(1-v_3v_4)}\ .
    \end{split}
\end{equation}
Note that in this case, we do not need. The giant graviton index is written as 
\begin{align}
    &\T{\C{I}}_{(m_1,m_2,0,0)}\sim\int \prod_{i=1}^{m_1}du_i^{(1)}\prod_{j=1}^{m_2}du_j^{(2)}\\
    &\;\times\frac{\Gamma(u_{1}+\delta_3-\delta_1,\delta_2+\delta_3,\delta_2+\delta_4)\Gamma(u_{1}+\tau,\delta_2+\delta_3,\delta_3+\delta_4)\Gamma(u_{1}+\sigma,\delta_2+\delta_3,\delta_3+\delta_4)}{\Gamma(u_{1},\delta_2+\delta_3,\delta_3+\delta_4)}\nonumber\\
    &\;\times\frac{\Gamma(u_{2}+\delta_4-\delta_2,\delta_3+\delta_4,\delta_1+\delta_4)\Gamma(u_{2}+\tau,\delta_3+\delta_4,\delta_1+\delta_4)\Gamma(u_{2}+\sigma,\delta_3+\delta_4,\delta_1+\delta_4)}{\Gamma(u_{2},\delta_3+\delta_4,\delta_1+\delta_4)}\nonumber\\
    &\;\times\frac{\theta(u_{12}+\frac{\delta_3+\delta_4}{2}-\frac{\tau-\sigma+1}{2},\delta_3+\delta_4)\theta(u_{21}+\frac{\delta_3+\delta_4}{2}-\frac{\tau-\sigma+1}{2};\delta_3+\delta_4)}{\theta(u_{12}+\frac{\delta_3+\delta_4}{2}-\frac{\tau+\sigma+1}{2},\delta_3+\delta_4)\theta(u_{21}+\frac{\delta_3+\delta_4}{2}-\frac{\tau+\sigma+1}{2};\delta_3+\delta_4)}\ ,\nonumber
\end{align}
with $\delta_1+\delta_2+\delta_3+\delta_4-\tau-\sigma=-1$. In large $N$ limit, this $4d$-$2d$ coupled matrix integral can be solved by the following two uniformly distributed parallelogram ansatz for each node:
\begin{align}
    &u_1=x_1(\delta_2+\delta_3)+y_1(\delta_3+\delta_4),\qquad u_2=x_2(\delta_1+\delta_4)+y_2(\delta_3+\delta_4)\ ,
\end{align}
where $-\frac{1}{2}<x_1,x_2,y_1,y_2<\frac{1}{2}$. From the $4d$ contributions, the prefactor in \eqref{eqn: SL3Z} is given by 
\begin{align}
    -\pi i m_1^2\frac{(\delta_3-\delta_1)\sigma\tau}{(\delta_2+\delta_3)(\delta_3+\delta_4)}
    -\pi i m_2^2\frac{(\delta_4-\delta_2)\sigma\tau}{(\delta_1+\delta_4)(\delta_3+\delta_4)}\ ,
\end{align}
Similarly, the $2d$ contribution from the bi-fundamental hypermultiplet can be obtained by the $SL(2,\mathbb{Z})$ transformation of $\theta$ function \eqref{eqn: SL(2,Z)}. In this case, the prefactor $B(z,\tau)$ is simply given by
\begin{align}
    2\pi i m_1 m_2\frac{\sigma\tau}{\delta_3+\delta_4}
\end{align}
The GG-expansion becomes
\begin{align}
    \frac{\C{I}_{T^{1,1}[N]}}{\C{I}_{T^{1,1}[\infty]}}
    &\sim \sum_{m_1,m_2=0}^{\infty} e^{2\pi i (\delta_1 m_1 +\delta_2 m_2) N}e^{-\pi i m_1^2\frac{(\delta_3-\delta_1)\sigma\tau}{(\delta_2+\delta_3)(\delta_3+\delta_4)}
    -\pi i m_2^2\frac{(\delta_4-\delta_2)\sigma\tau}{(\delta_1+\delta_4)(\delta_3+\delta_4)}{\delta_2\delta_3}+2\pi i m_1 m_2\frac{\sigma\tau}{\delta_3+\delta_4}}\nonumber\\
    &\sim \int_0^\infty dm_1dm_2\; \exp\bigg[2\pi i (\delta_1 m_1 +\delta_2 m_2) N-\pi i m_1^2\frac{(\delta_3-\delta_1)\sigma\tau}{(\delta_2+\delta_3)(\delta_3+\delta_4)}
    \\
    &\hspace{4cm}-\pi i m_2^2\frac{(\delta_4-\delta_2)\sigma\tau}{(\delta_1+\delta_4)(\delta_3+\delta_4)}{\delta_2\delta_3}+2\pi i m_1 m_2\frac{\sigma\tau}{\delta_3+\delta_4}\bigg]\ .\nonumber
\end{align}
By extremizing the integral with respect to the wrapping number $m_1$ and $m_2$, we obtain
\begin{equation}
    \log \C{I}_{T^{1,1}[N]} = -\pi i N^2\frac{\delta_1\delta_2\delta_3+\delta_2\delta_3\delta_4+\delta_1\delta_2\delta_4+ \delta_1\delta_3\delta_4}{\sigma \tau}\ .
\end{equation}

\paragraph{Small black hole} Similar to the $\C{N}=4$ $U(N)$ SYM, one can consider the small black hole dual to $T^{1,1}$. We set the partially equal charges $Q_1=Q_3$ and $Q_2=Q_4$, or, equivalently, $\delta_1=\delta_3$ and $\delta_2=\delta_4$ where the small black hole limit is characterized by $\delta_1,\delta_2\ll 1$. Firstly, we shall show that $m_{1*}$ and $m_{2*}$ are real, which enables us to elucidate the entropy of the small black hole as a bound state of D3-branes through the giant graviton index $\T{\C{I}}_{(m_{1*},m_{2*},0,0)}$ with well-defined wrapping numbers

Since the charges are approximated as follows,
\begin{equation}
    \begin{split}
        Q_1\approx -\frac{N^2}{2}\frac{\delta_1\delta_2+2\delta_2^2}{\sigma\tau},\qquad Q_2\approx -\frac{N^2}{2}\frac{\delta_1\delta_2+2\delta_1^2}{\sigma\tau}\ .
    \end{split}
\end{equation}
The extremized wrapping numbers are given by
$m_{1*}=-N\frac{(\delta_1+\delta_2)\delta_2}{\sigma\tau}$ and $m_{2*}=-N\frac{(\delta_1+\delta_2)\delta_1}{\sigma\tau}$.
One can rewrite $m_{1*}$ and $m_{2*}$ in terms of charges:
\begin{align}
    \begin{split}
        m_{1*}&\cong\frac{1}{N}\left(Q_1-\frac{N^2}{2}\frac{\delta_1\delta_2}{\sigma\tau}\right)\cong\frac{1}{N}\left(Q_1+\frac{1}{3}(Q_1+Q_2-\sqrt{Q_1^2-Q_1Q_2+Q_2^2})\right)\ ,\\
        m_{2*}&\cong\frac{1}{N}\left(Q_2-\frac{N^2}{2}\frac{\delta_1\delta_2}{\sigma\tau}\right)\cong\frac{1}{N}\left(Q_2+\frac{1}{3}(Q_1+Q_2-\sqrt{Q_1^2-Q_1Q_2+Q_2^2})\right)\ .
    \end{split}
\end{align}
Therefore, $m_{1*}$ and $m_{2*}$ are real because $Q_1$ and $Q_2$ are real.

The single-letter index of the adjoint fields becomes trivial in the small black hole limit, while the single-letter index of the bi-fundamental fields is written as
\begin{align}
    f_{12}=\sqrt{\frac{v_1 v_2}{pq}}\frac{(1-p)(1-q)}{(1-v_1 v_2)}\ .
\end{align}
which is that of $2d$ $\C{N}=(2,2)$ bifundamental hypermultiplets. Thus, the entropy of the effective $2d$ system is given by extremizing the following Legendre transformation ($\sigma,\tau\to\frac{1}{2}$)
\begin{equation}
    S(Q)=-8\pi i N^2\delta_1\delta_2(\delta_1+\delta_2) -4\pi i (\delta_1 Q_1+\delta_2 Q_2)\ ,
\end{equation}
and taking the real part, which leads us to
\begin{align}
    S\cong\frac{4 \sqrt{2} \pi  \left(\sqrt{Q_1^2-Q_1 Q_2+Q_2^2}+2
   Q_1-Q_2\right) \sqrt{2 \sqrt{Q_1^2-Q_1 Q_2+Q_2^2}-2
   Q_1+Q_2}}{3\sqrt{3} N}\ .
\end{align}
For $Q_1=Q_2=Q$, the entropy is further simplified as $S\approx \frac{8\sqrt{2}\pi Q^{3/2}}{3\sqrt{3}N}$.

\subsubsection{$Y^{p,q}$}
Let us now consider an infinite family of $\C{N}=1$ toric quiver gauge theories, denoted as $Y^{p,q}$, where $p>q$ are positive integers. Similar to the case of $T^{1,1}$, an appropriate choice of fugacity degrees allows us to reduce the multiple-sum to a double-sum. The GG-expansion is written as
\begin{align}
    \frac{\C{I}_{N}}{\C{I}_{\infty}}=\sum_{m_1,m_2=0}^\infty v_1^{m_1N}v_2^{m_2N} \T{\C{I}}_{(m_1,m_2,0,0)}
\end{align}
Let us first fix the order of projection $k_I$ and fugacities $w_r$. For $Y^{p,q}$, there are four corners labelled by $I=1,2,3,4$ and edges labelled by $r \in \{1/2, 1+1/2, 2+1/2, 3+1/2\}$. The primitive vectors $v_I$ in the toric diagram are given by
\begin{equation}
    (x_I, y_I) = \{(0,0), (1,0), (p,p), (p-q-1,p-q)\}\ .
\end{equation}
Explicitly, the $g_r$ for $r \in \{1/2, 1+1/2, 2+1/2, 3+1/2\}$ are:
\begin{equation}
    \begin{split}
        g_{1/2} &= (p-q, 1-p+q, 0), \\
        g_{1+1/2} &= (0,1,0), \\
        g_{2+1/2} &= (-p, -1+p, p), \\
        g_{3+1/2} &= (q, -1-q, p)\ .
    \end{split}
\end{equation}
which leads
\begin{equation}
    \begin{split}
        w_{1/2} &= v_2^{p-q}v_3^{p}, \\
        w_{1+1/2} &= v_3^{p}v_4^{p-q}, \\
        w_{2+1/2} &= v_4^{p+q}v_1^{p}, \\
        w_{3+1/2} &= v_1^{p}v_2^{p+q}\ .
    \end{split}
\end{equation}
Finally, the degeneracy values $k_I$ are then determined by the area of the parallelogram made by two edges $I-\frac{1}{2}$ and $I+\frac{1}{2}$:
\begin{align}
    k_1=p-q,\qquad k_2=p, \qquad k_3=p+q, \qquad k_4=p\ .
\end{align}
In other words, the $4d$-$2d$ system of giant graviton is given by a quiver theory, which consists of $Z_p$ and $Z_{p-q}$ orbifold with $2d$ bifundamental fields.

The free energy can be obtained by extremizing the following function with $m_1$ and $m_2$ as
\begin{equation}
    \begin{split}
        \log\C{I}_{Y^{p,q}}\sim&-\pi i \frac{1}{p}(p m_2)^2\frac{(\delta_4-\delta_2)\sigma\tau}{\big(\delta_3+\frac{p-q}{p}\delta_4\big)\big(\delta_1+\frac{p+q}{p}\delta_4\big)}\\
        &-\pi i \frac{1}{p-q}\big((p-q) m_1\big)^2\frac{\left(\frac{p+q}{p-q}\delta_3-\delta_1\right)\sigma\tau}{\big(\delta_2+\frac{p}{p-q}\delta_3\big)\big(\delta_4+\frac{p}{p-q}\delta_3\big)}\\
        &+2\pi i (p-q) m_1 N\delta_1+2\pi i (p m_2) N\delta_2\\
        &+2\pi i p (p-q)m_1 m_2 \frac{\sigma\tau}{p\delta_3+(p-q)\delta_4}\ .
    \end{split}
\end{equation}
which reproduces the desired answer:
\begin{equation}
    \log\C{I}_{Y^{p,q}}\sim-i\pi N^2\frac{p\delta_1\delta_2\delta_3+(p+q)\delta_1\delta_2\delta_4+p\delta_1\delta_3\delta_4+(p-q)\delta_2\delta_3\delta_4}{\sigma\tau}.
\end{equation}

\section{Conclusion}\label{sec: conclusion}
In this paper, we revisit the microstate counting problem of BPS black holes in AdS from the gravity perspective using the GG-expansion. 
Our study provides compelling evidence supporting the validity of the GG-expansion of the index; it accounts for the black hole entropy with $O(N^2)$ degrees of freedom from giant gravitons with $O(N)$ degrees of freedom. We find that the universal large $N$ saddle, called parallelogram ansatz, solves the matrix integral realized on the wrapped D-branes for a wide range of parameter space. 
Furthermore, we discover that the fugacity condition for the ansatz (with equal electric chemical potentials) to be valid is the same as the CTC-free and stability condition of the black hole. 
However, the physical reasons behind their equivalence remain elusive. 

We would like to emphasize that in our computations, simplifying multiple-sum into the simple-sum of the GG-expansion played a significant role. This reduction enables us to test the GG-expansion across various holographic examples, including the insertion of surface defects, orbifolds, and toric quiver gauge theories.

We also provide a concrete realization of the D-brane based descriptions for the dual small black holes in $\C{N}=4$ SYM and $T^{1,1}$. In particular, in certain chemical potential regimes, the system can be characterized by an effective $2d$ theory, which reminds us of the concept from Kerr/CFT$_2$ correspondence \cite{Guica:2008mu}. 
According to Kerr/CFT$_2$ correspondence, the near-horizon geometry of extremal black holes exhibits asymptotic symmetries, with generators satisfying the Virasoro algebra \cite{Chow:2008dp}. 
Therefore, it is tempting to explore the connection between the effective 2D description and the Virasoro algebra in the context of Kerr/CFT$_2$ correspondence.

We examined the half-BPS defect, which can be regarded as a probe serving as an order parameter for the deconfinement phase transition. However, if the defect is sufficiently heavy, such that the defect central charge scales as $O(N^2)$, the gravitational backreaction of probing D3-branes in the bulk side cannot be ignored, leading to the emergence of a bubbling geometry \cite{Lin:2004nb,Gomis:2007fi}. Analyzing the bubbling geometries and their phases in terms of the GG-expansion would be an interesting avenue for future research.

In addition, one might explore various extensions of our work by considering the GG-expansion of holographic field theories with different dimensions. 
For instance, superconformal field theory on an $M2$ brane is dual to string theory in $AdS_4 \times S^7$, where the giant graviton expansion of the superconformal index involves M5 branes wrapping $S^5 \subset S^7$. Similarly, superconformal field theory on an $M2$ brane is dual to string theory in $AdS_7 \times S^4$, with the giant graviton expansion of the superconformal index involving M2 branes wrapping $S^2 \subset S^4$ \cite{Arai:2020uwd,Gaiotto:2021xce,Choi:2022ovw}. 
Thus, a potential relationship may exist between the index of $6d/3d$ SCFTs on M5/M2-branes via the GG-expansion, shedding light on formulating the $6d$ index. 
Furthermore, it has been found that the holographic dual of Argyres-Douglas theories, which are non-Lagrangian, strongly coupled, and interacting theories, are the so-called spindle geometries \cite{Bah:2021hei, Bah:2021mzw}.
Hence, it may be possible to test the GG-expansion even in these highly non-trivial examples, where the precise form of the index is not well-known.

Finally, the GG-expansion formalism can help us understand the spectrum of field theory from the gravity perspective. 
For example, operators representing black hole states have been discovered at low ranks of $\mathcal{N}=4$ $U(N)$ SYM \cite{Chang:2022mjp,Choi:2022caq,Choi:2023znd, Choi:2023vdm, Chang:2023zqk, Budzik:2023vtr}. 
In particular, the GG-expansion provides valuable insights for identifying operators describing small black holes in AdS, which locally resemble flat space. 
A potential interpretation of these black hole operators within this framework suggests that they resemble determinant-like operators with open strings ending on them. 

\begin{acknowledgments}
We thank Minseok Cho, Sunjin Choi, Seok Kim, and Ki-Hong Lee for helpful discussions. We are especially grateful to Jaewon Song for many valuable feedback on the draft. This work is supported by the National Research Foundation of Korea (NRF) Grant RS-2023-00208602 (SK) and 2021R1A2C2012350 (EL).
\end{acknowledgments}

\appendix
\section{\texorpdfstring{Parellelogram ansatz}{}}\label{sec: appendix A}

In this Appendix, we prove that the ansatz \eqref{eqn: Ansatz} solves the saddle point equation, which accounts for the free energy of the dual black hole \eqref{eqn: BH entropy}. For simplicity, let us choose the branch $\delta_1+\delta_2+\delta_3-\sigma-\tau=-1$. Applying modular identity, with the prefactor $P_{+}$, to \eqref{eqn: N=4 SYM (3)} yields
\begin{align}
    Z=\exp\left[-\frac{\pi i N^2 \delta_1\delta_2\delta_3}{\sigma\tau}\right]\frac{1}{N!}\exp\left[-\sum_{a\neq b}\left(V_{\sigma}(u_{ab})+\tilde{V}_{\tau}(u_{ab})\right)\right]
\end{align}
where
\begin{eqnarray}\label{potential-1}
  -V_\sigma(u)&\equiv&\frac{1}{2}\sum_{I=1}^3\log\Gamma({\textstyle
  -\frac{\delta_I+u+1}{\sigma},-\frac{1}{\sigma},-\frac{\tau}{\sigma}})
  -\frac{1}{2}\log\Gamma({\textstyle -\frac{u+1}{\sigma},-\frac{1}{\sigma},
  -\frac{\tau}{\sigma}})+\left(u\rightarrow-u\right),
  \nonumber\\
  -\tilde{V}_\tau(u)&\equiv&\frac{1}{2}\sum_{I=1}^3\log\Gamma({\textstyle
  \frac{\delta_I+u}{\tau},-\frac{1}{\tau},\frac{\sigma}{\tau}})
  -\frac{1}{2}\log\Gamma({\textstyle \frac{u}{\tau},-\frac{1}{\tau},
  \frac{\sigma}{\tau}})+\left(u\rightarrow-u\right).\
\end{eqnarray}
The integral can be rewritten as follows:
\begin{equation}\label{eqn: YM-final}
  Z=\exp\left[-\frac{\pi iN^2\delta_1\delta_2\delta_3}{\sigma\tau}\right]
  \int\prod_{a=1}^N du_a\cdot\frac{1}{N!}
  \prod_{a\neq b}(1-e^{\frac{2\pi iu_{ab}}{\tau}})\cdot
  \exp\left[-\sum_{a\neq b}\left(V_\sigma(u_{ab})+V_\tau(u_{ab})\right)\right]\ ,
\end{equation}
where $V_{\tau}$ is given by
\begin{equation}\label{V-tilde-V}
  e^{-\sum_{a\neq b}\tilde{V}_\tau(u_{ab})}=
  \prod_{a\neq b}(1-e^{\frac{2\pi iu_{ab}}{\tau}})
  \cdot e^{-\sum_{a\neq b}V_\tau(u_{ab})}.\
\end{equation}
We use an integral identity called Molien-Weyl formula that can be applied for a permutation invariant function $f(u)$: 
\begin{align}
    \prod_{a=1}^N du_a\frac{1}{N!}\prod_{a\neq b}(1-e^{\frac{2\pi iu_{ab}}{\tau}}) f(u)
    =\prod_{a=1}^N du_a \prod_{a< b}(1-e^{\frac{2\pi iu_{ab}}{\tau}}) f(u).
\end{align}
Using the identity, the integral is written as
\begin{equation}\label{YM-final}
  Z=\exp\left[-\frac{\pi iN^2\delta_1\delta_2\delta_3}{\sigma\tau}\right]
  \int\prod_{a=1}^N du_a\cdot
  \prod_{a < b} (1-e^{\frac{2\pi iu_{ab}}{\tau}})
  \exp\left[-\sum_{a\neq b}\left(V_\sigma(u_{ab})+V_\tau(u_{ab})\right)\right].\
\end{equation}
Let us note that $V_{\sigma}$ is the periodic potential in the $\sigma$ direction, and $V_{\tau}$ is the periodic potential in the tau direction.
As we show below, by using the periodicity of these potentials, one can demonstrate that the saddle of the integral is governed by the uniform eigenvalue distribution on a parallelogram with two edges given as $\sigma$ and $\tau$:\footnote{Since we are considering the large $N$ limit, one can approximate the discrete eigenvalue distribution as the continuous density function $\rho(x,y)$.}
\begin{align}\label{eqn: ansatz appendix}
    u(x,y)=\sigma x+ \tau y,\quad \rho(x,y)=1,
\end{align}
where $\rho(x,y)$ is the areal density satisfying $\int dx dy \rho(x,y)=1$. We order the eigenvalues $x_a>x_b$ if $a<b$.

Let us verify whether the ansatz satisfies the saddle point equation.
The force-free condition for an eigenvalue $u_{a}$ is written as
\begin{align}\label{eqn: force-free N=4}
    \int_{-1/2}^{1/2} dx_b dy_b\left[\frac{1}{\sigma}\frac{\partial}{\partial x_a}V_{\sigma}(u_{ab})+\frac{1}{\tau}\frac{\partial}{\partial y_a}V_{\tau}(u_{ab})\right] = 0.
\end{align}
When $V_{\sigma}(u_{ab})$ and $V_{\tau}(u_{ab})$ are regular for all $u_{a},u_b$ inside the parallelogram, the equation can be reformulated as
\begin{align}
    \int_{-1/2}^{1/2} dy_b\frac{1}{\sigma}\left[V_{\sigma}(u_{ab})\big|_{x_b=-1/2}^{x_b=1/2}+\int_{-1/2}^{1/2} dx_a\frac{1}{\tau}V_{\tau}(u_{ab})\big|_{y_b=-1/2}^{y_b=1/2}\right] = 0.
\end{align}
Since $V_{\sigma}(u_{ab})\big|_{x=-1/2}^{x=1/2}$ and $V_{\tau}(u_{ab})\big|_{y=-1/2}^{y=1/2}$ both vanish due to the periodicity, it solves the saddle point equation.
Therefore, we prove that the free energy of this ansatz is given as $\log Z = -\pi i N^2\frac{\delta_1\delta_2\delta_3}{\sigma\tau}$ of which the Legendre transformation reproduces the entropy of BPS black holes.

One can be convinced that the conditions for 
$V_{\sigma}(u_{ab})$ and $V_{\tau}(u_{ab})$ to be regular, \textit{i.e.} no branch points/cuts, for the ansatz\eqref{eqn: ansatz appendix} are given as follows:
\begin{equation}\label{condition-intrinsic4}
  {\rm Im}\left(\frac{\sigma-\delta_I}{\tau}\right)<0\ ,\ \
  {\rm Im}\left(\frac{1+\delta_I}{\tau}\right)<0\ ,\ \
  {\rm Im}\left(\frac{1-\tau+\delta_I}{\sigma}\right)<0\ ,\ \
  {\rm Im}\left(\frac{\delta_I}{\sigma}\right)>0\ .
\end{equation}
For $\delta_1+\delta_2+\delta_3-\sigma-\tau=+1$, on the other hand, one can apply modular identity involving $P_{-}$.
The free energy appears to be the same.
The parallelogram ansatz becomes a saddle if
\begin{equation}
  {\rm Im}\left(\frac{\delta_I-\tau}{\sigma}\right)<0\ ,\ \
  {\rm Im}\left(\frac{1-\delta_I}{\sigma}\right)<0\ ,\ \
  {\rm Im}\left(\frac{1+\sigma-\delta_I}{\tau}\right)<0\ ,\ \
  {\rm Im}\left(\frac{\delta_I}{\tau}\right)<0\ .
\end{equation}
where we assumed $\text{Im}\frac{\sigma}{\tau}>0$ without losing generality.

\section{Derivation of matrix integral on D3-branes wrapped around $S^3/\mathbb{Z}_k$}\label{sec: appendix B}
In this appendix, we derive the matrix integral \eqref{eqn: Orbifold GGE} realized on the giant graviton wrapped around $S^3/\mathbb{Z}_k\subset S^5$ and do the saddle point approximation using the parallelogram ansatz in large $N$ limit. Let us begin by computing the single-letter index
\begin{equation}
    \T{\C{I}}_{U(\B{m})^k}=\int_{U(\B{m})^k}dg\;\textrm{PE}\left[\sigma_1 P_k\left( f_v(p,q,x,y,z)\Tr U^\dagger\Tr U\right)\right]
\end{equation}
where
\begin{equation}
    \begin{split}
        \sigma_1 P_k\bigg( f_v&(p,q,x,y,z)\Tr U^\dagger\Tr U\bigg)\\
        &=\frac{1}{k}\sum_{l=0}^{k-1}\bigg(1-\frac{(1-x^{-1}\omega^{-\alpha_1 l})(1-p)(1-q)}{(1-y \omega^{\alpha_2 l})(1-z \omega^{\alpha_3 l})}\sum_{a,b=1}^{k}\Tr U_a^{\dagger}\Tr U_b \omega^{(a-b)l}\bigg)
    \end{split}
\end{equation}
For the first part of the right-hand-side of the equation, we obtain
\begin{equation}
    \frac{1}{k}\sum_{l=0}^{k-1}\sum_{a,b=1}^{k}\Tr U_a^{\dagger}\Tr U_b \omega^{(a-b)l}=\sum_{a=1}^k\Tr U_a^\dagger \Tr U_a
\end{equation}
This will cancel, except for the diagonal part, with the Haar measure from $\int_{U(\B{m})^k}dg$. For the remaining terms, we employ the following trick
\begin{align}
    -&\frac{1}{k}\sum_{l=0}^{k-1}\frac{(1-x^{-1}\omega^{-\alpha_1 l})(1-p)(1-q)}{(1-y \omega^{\alpha_2 l})(1-z \omega^{\alpha_3 l})}\sum_{a,b=1}^{k}\Tr U_a^{\dagger}\Tr U_b \omega^{(a-b)l}\\
    &=-\frac{1}{k}\sum_{a,b=1}^k\sum_{l=0}^{k-1}\frac{(1-x^{-1}\omega^{-\alpha_1 l})(1-p)(1-q)}{(1-y \omega^{\alpha_2 l})(1-z \omega^{\alpha_3 l})}\frac{\sum_{m=0}^{k-1}(y\omega^{\alpha_2l})^{m}\sum_{n=0}^{k-1}(z\omega^{\alpha_3 l})^{n}}{\sum_{m=0}^{k-1}(y\omega^{\alpha_2l})^{m}\sum_{n=0}^{k-1}(z\omega^{\alpha_3 l})^{n}}\;\Tr U_a^{\dagger}\Tr U_b\;\omega^{(a-b)l}\nonumber\\
    &=-\frac{1}{k}\sum_{a,b=1}^k\sum_{l,m,n=0}^{k-1}\frac{(1-x^{-1}\omega^{-\alpha_1 l})(1-p)(1-q)(y\omega^{\alpha_2l})^{m}(z\omega^{\alpha_3 l})^{n}}{(1-y^k)(1-z^k)}\omega^{(a-b)l}\Tr U_a^{\dagger}\Tr U_b \nonumber\\
    &=-\frac{1}{k}\sum_{a,b=1}^k\sum_{l,m,n=0}^{k-1}\frac{(1-p)(1-q)}{(1-y^k)(1-z^k)}\big[y^m z^n\omega^{(\alpha_2m+\alpha_3n+a-b)l}\nonumber\\
    &\hspace{5cm}-x^{-1}y^{m}z^{n}\omega^{(-\alpha_1+\alpha_2m+\alpha_3n+a-b)l}\big]\Tr U_a^{\dagger}\Tr U_b\nonumber
\end{align}
Performing summation over $l$ yields
\begin{equation}
    -\frac{1}{k}\cdot k\sum_{a=1}^k\sum_{m,n=0}^{k-1}\frac{(1-p)(1-q)}{(1-y^k)(1-z^k)}\big(y^{m}z^{n}\Tr U_a^{\dagger}\Tr U_{b_*}-x^{-1}y^{m}z^{n}\Tr U_a^{\dagger}\Tr U_{b_{**}}\big)
\end{equation}
where
\begin{equation}
    b_*(a)=\alpha_2 m+\alpha_3n+a,\qquad b_{**}(a)=-\alpha_1+\alpha_2m+\alpha_3n+a,\qquad \mod k
\end{equation}
We can rewrite the second term as follows
\begin{align}
    \begin{split}
        \sum_{a=1}^k\sum_{m,n=0}^{k-1}&x^{-1}y^m z^n \Tr U_a^\dagger\Tr U_{b_{**}(a)}\\
        &=\sum_{b_*=1}^k\sum_{m,n=0}^{k-1}x^{-1}y^{k-1-m}z^{k-1-n}\Tr U_{b_*}^\dagger\Tr U_{b_{**}(b_*)}
    \end{split}
\end{align}
since
\begin{align}
    b_*(a)&=\alpha_2m+\alpha_3n+a,\quad \mod k\\
    b_{**}(b_*(a))&=-\alpha_1+\alpha_2(k-1-m)+\alpha_3(k-1-n)+\alpha_2m+\alpha_3n+a=a,\quad \mod k \nonumber
\end{align}
Altogether, we get
\begin{equation}\label{eqn: letter}
    -\frac{(1-p)(1-q)}{(1-y^k)(1-z^k)}\sum_{a=1}^k\sum_{m,n=0}^{k-1}\big[(y^mz^n\Tr U_a^\dagger\Tr U_{b_*(a)}-x^{-1}y^{k-1-m}z^{k-1-n}\Tr U_{b_*(a)}^\dagger\Tr U_a\big]
\end{equation}
To proceed, we note that the elliptic gamma function can be defined by
\begin{equation}
    \Gamma(A,B,C)=\textrm{PE}\left[\frac{A-\frac{BC}{A}}{(1-B)(1-C)}\right]
\end{equation}
The key observation is that in numerator the product of two terms is equal to $BC$. Now, the expression \eqref{eqn: letter} has the following structure
\begin{equation}
    \begin{split}
        \frac{pqy^mz^n-\frac{y^{k-1-m}z^{k-1-m}}{x}}{(1-y^k)(1-z^k)},&\quad \frac{y^mz^n-pq\frac{y^{k-1-m}z^{k-1-m}}{x}}{(1-y^k)(1-z^k)}\\
        \frac{qy^mz^n-p\frac{y^{k-1-m}z^{k-1-m}}{x}}{(1-y^k)(1-z^k)},&\quad \frac{py^mz^n-q\frac{y^{k-1-m}z^{k-1-m}}{x}}{(1-y^k)(1-z^k)}
    \end{split}
\end{equation}
Using $pq=xyz$, we see that in all cases, the product of two terms in the numerator is given by $y^kz^k$. This means that each combination can be correctly packaged into an elliptic gamma function. Therefore, it can be read off
\begin{equation}
    \begin{split}
        &\T{\C{I}}_{U(\B{m})^k}\sim\oint\prod_{a=1}^k\prod_{i=1}^{\B{m}}du_i^{(a)}\prod_{b,c=0}^{k-1}\prod_{i,j}^{\B{m}}\frac{\Gamma(u_{ij}^{(a, [\![ a+b\alpha_2+c\alpha_3 ]\!])}+b\delta_2+c\delta_3+\sigma, k\delta_2,k\delta_3)}{\Gamma(u_{ij}^{(a, [\![ a+b\alpha_2+c\alpha_3 ]\!])}+b\delta_2+c\delta_3, k\delta_2,k\delta_3)}\\
        &\hspace{4.8cm}\times \frac{\Gamma(u_{ij}^{(a, [\![ a+b\alpha_2+c\alpha_3 ]\!])}+b\delta_2+c\delta_3+\tau, k\delta_2,k\delta_3)}{\Gamma(u_{ij}^{(a, [\![ a+b\alpha_2+c\alpha_3 ]\!])}+b\delta_2+c\delta_3+\tau+\sigma, k\delta_2,k\delta_3)}
    \end{split}
\end{equation}
where for an integer $l$ we defined $ [\![ l ]\!]$ to be modulo $k$ and ignored some overall factors.

To evaluate this matrix integral in large $N$ limit using the saddle point approximation, we reformulate the integrand using the modular property of $SL(3,\mathbb{Z})$ \eqref{eqn: SL3Z} with the modular parameters $\delta_2$ and $\delta_3$ on giant graviton:
\begin{equation}
    \C{I}_{U(\B{m})^k}\sim\oint\prod_{a=1}^k\prod_{i=1}^{\B{m}}du_i^{(a)}\;e^{-i\pi\bm{P}_+}e^{\sum_{b,c=0}^{k-1}\big(-V_{\delta_2}-V_{\delta_3}\big)}
\end{equation}
where we omitted a possible overall normalization factor, which is irrelevant in the following discussion. According to \cite{Choi:2023tiq}, the prefactor is completely governed by the trace anomalies, and only captures the 't Hooft anomalies in a consistent theory without gauge anomalies, which must be also true for the theory on giant gravitons:
\begin{equation}
    -i\pi\bm{P}_+=-i\pi k\B{m}^2\frac{(-\delta_1)\sigma\tau}{\delta_2\delta_3}
\end{equation}
On the other hand, the potential terms are given by
\begin{equation}
    -V_{\delta_2}=-v_{\delta_2}(u_{ij}^{(a,[\![ a+b\alpha_2+c\alpha_3 ]\!]}+b\delta_2+c\delta_3),\quad-V_{\delta_3}=-v_{\delta_3}(u_{ij}^{(a,[\![ a+b\alpha_2+c\alpha_3 ]\!]}+b\delta_2+c\delta_3)\ ,
\end{equation}
where
\begin{align}
    -&v_{\delta_2}(u)\\
    &\equiv\sum_{i,j}\log\Gamma\left(-\frac{u_{ij}^{(a,a^{\prime})}+\sigma+1}{k\delta_2},-\frac{1}{k\delta_2},-\frac{\delta_3}{\delta_2}\right)+\sum_{i,j}\log\Gamma\left(-\frac{u_{ij}^{(a,a^{\prime})}+\tau+1}{k\delta_2},-\frac{1}{k\delta_2},-\frac{\delta_3}{\delta_2}\right)\nonumber\\
    &-\sum_{i,j}\log\Gamma\left(-\frac{u_{ij}^{(a,a^{\prime})}+1}{k\delta_2},-\frac{1}{k\delta_2},-\frac{\delta_3}{\delta_2}\right)-\sum_{i,j}\log\Gamma\left(-\frac{u_{ij}^{(a,a^{\prime})}+\sigma+\tau+1}{k\delta_2},-\frac{1}{k\delta_2},-\frac{\delta_3}{\delta_2}\right)\nonumber\ ,
\end{align}
and
\begin{align}
    -&v_{\delta_3}(u)\\
    &\equiv\sum_{i,j}\log\Gamma\left(-\frac{u_{ij}^{(a,a^{\prime})}+\sigma}{k\delta_3},-\frac{1}{k\delta_3},\frac{\delta_2}{\delta_3}\right)+\sum_{i,j}\log\Gamma\left(-\frac{u_{ij}^{(a,a^{\prime})}+\tau}{k\delta_3},-\frac{1}{k\delta_3},\frac{\delta_2}{\delta_3}\right)\nonumber\\
    &-\sum_{i,j}\log\Gamma\left(-\frac{u_{ij}^{(a,a^{\prime})}}{k\delta_3},-\frac{1}{k\delta_3},\frac{\delta_2}{\delta_3}\right)-\sum_{i,j}\log\Gamma\left(-\frac{u_{ij}^{(a,a^{\prime})}+\sigma+\tau}{k\delta_3},-\frac{1}{k\delta_3},\frac{\delta_2}{\delta_3}\right)\nonumber\ .
\end{align}

Now, we introduce our uniform parallelogram ansatz in large $N$ limit. Since we are only concerned with $U(\B{m})$ gauge group for all gauge nodes, the ansatz is given by
\begin{equation}
    u^{(a)}=\delta_2 x+\delta_3 y,\qquad x,y\in\left(-\frac{1}{2},\frac{1}{2}\right)
\end{equation}
with a $2$-dimensional eigenvalue density $\rho^{(a)}(x,y)$, which satisfies $\int dxdy\rho^{(a)}(x,y)=1$. To demonstrate that the parallelogram ansatz indeed represents large $N$ saddle points, we need to confirm that the potentials $V_{\delta_2}(u)$ and $V_{\delta_3}(u)$ have vanishing forces at the leading order in $N$ within the parallelogram region. Taking a derivative of the potential with respect to a particular eigenvalue $u_i^{(a)}$, we compute the force acting on $u_i^{(a)}$:
\begin{align}
    &\sum_{b,c=0}^{k-1}\sum_j\bigg[\frac{\partial}{\partial u_i^{(a)}}V_{\delta_2}+\frac{\partial}{\partial u_i^{(a)}}V_{\delta_3}\bigg]\\
    &\;\sim\sum_{b,c=0}^{k-1}\sum_j\bigg[\frac{\partial}{\partial u_i^{(a)}}v_{\delta_2}\big(u_{ij}^{(a,[\![ a+b\alpha_2+c\alpha_3 ]\!])}+b\delta_2+c\delta_3\big)+\frac{\partial}{\partial u_i^{(a)}}v_{\delta_3}\big(u_{ij}^{(a,[\![ a+b\alpha_2+c\alpha_3 ]\!])}+b\delta_2+c\delta_3\big)\bigg]\nonumber\\
    &\;\sim\sum_{b,c=0}^{k-1}\sum_j\bigg[\frac{1}{\delta_2}\frac{\partial}{\partial x_j}v_{\delta_2}(u_{ij}^{(a,[\![ a+b\alpha_2+c\alpha_3 ]\!])}+b\delta_2+c\delta_3)+\frac{1}{\delta_3}\frac{\partial}{\partial y_j}v_{\delta_3}(u_{ij}^{(a,[\![ a+b\alpha_2+c\alpha_3 ]\!])}+b\delta_2+c\delta_3)\bigg]\nonumber\ ,
\end{align}
where we used $\frac{\partial}{\partial u_i^{(a)}}  \sim\frac{\partial}{\partial u_j^{(a^\prime)}}$ and $u_j^{(a^\prime)}=\delta_2 x_j+\delta_3 y_j$ so that the derivatives can be replaced by either $\frac{1}{\delta_2}\frac{\partial}{\partial x_j}$ or $\frac{1}{\delta_3}\frac{\partial}{\partial y_j}$. In the large $N$ continuum limit, the above force becomes
\begin{align}
    &\B{m}\sum_{b,c=0}^{k-1}\int_{-1/2}^{1/2}\int_{-1/2}^{1/2}dx_2dy_2\left(\frac{1}{\delta_2}\frac{\partial}{\partial x_2}v_{\delta_2}(\delta_2 x_{12} +\delta_3 y_{12})+\frac{1}{\delta_3}\frac{\partial}{\partial y_2}v_{\delta_3}(\delta_2 x_{12} +\delta_3 y_{12})\right) \\
    &\;=\frac{\B{m}}{\delta_2}\int_{-1/2}^{k-1/2}dy_2\; v_{\delta_2}(\delta_2 x_{12}+\delta_3 y_{12})\bigg|^{x_2 = k-1/2}_{x_2 = -1/2} +\frac{\B{m}}{\delta_3}\int_{k-1/2}^{1/2}dx\;v_{\delta_3}(\delta_2x_{12}+\delta_3y_{12})\bigg|^{y = k-1/2}_{y = -1/2}\nonumber\ ,
\end{align}
where $x_{12}=x_1-x_2$ and $y_{12}=y_1-y_2$, and we used
\begin{equation}
    \int_{-1/2}^{1/2}dx \left[f\left(\frac{x}{k}\right)+f\left(\frac{x}{k}+\frac{1}{k}\right)+\cdots+f\left(\frac{x}{k}+\frac{k-1}{k}\right)\right]=\int_{-1/2}^{k-1/2}f\left(\frac{x}{k}\right)dx
\end{equation}
We see that both terms vanish when $v_{\delta_2}$ and $v_{\delta_3}$ are periodic in $k\delta_2$ and $k\delta_3$ respectively. Namely, the force-free condition is satisfied if
\begin{equation}\label{eqn: force-free condition}
        v_\sigma( u+ k\delta_2) = v_\sigma(u)\ , \quad v_{\delta_3}( u+ k\delta_3) = v_\tau(u) \ .
\end{equation}
Indeed, this is true because of the periodicity of the elliptic gamma function $\Gamma(A,B,C)=\Gamma(A+1,B,C)$. However, this is not the end of the story. As already explained in Appendix~\ref{sec: appendix A}, we need to check whether there exist branch points/cuts inside the parallelogram region to avoid disruption of the periodicity of $v_{\delta_2}$ and $v_{\delta_\tau}$. One can obtain the conditions along similar lines:
\begin{equation}
    \Im\left(\frac{\delta_I^\prime-k\delta_3}{k\delta_2}\right)<0,\quad \Im\left(\frac{1-\delta_I^\prime}{k\delta_2}\right)<0,\quad \Im\left(\frac{1+k\delta_2-\delta_I^\prime}{k\delta_3}\right)<0,\quad \Im\left(\frac{\delta_I^\prime}{k\delta_3}\right)<0
\end{equation}
where $\delta_I^\prime=(\sigma,\tau,-\sigma-\tau)$.



\bibliographystyle{JHEP}
\bibliography{ref}

\providecommand{\href}[2]{#2}\begingroup\raggedright\begin{thebibliography}{10}

\bibitem{Strominger:1996sh}
A.~Strominger and C.~Vafa, {\it {Microscopic origin of the Bekenstein-Hawking entropy}},  {\em Phys. Lett. B} {\bf 379} (1996) 99--104, [\href{http://arxiv.org/abs/hep-th/9601029}{{\tt hep-th/9601029}}].

\bibitem{Gutowski:2004ez}
J.~B. Gutowski and H.~S. Reall, {\it {Supersymmetric AdS(5) black holes}},  {\em JHEP} {\bf 02} (2004) 006, [\href{http://arxiv.org/abs/hep-th/0401042}{{\tt hep-th/0401042}}].

\bibitem{Gutowski:2004yv}
J.~B. Gutowski and H.~S. Reall, {\it {General supersymmetric AdS(5) black holes}},  {\em JHEP} {\bf 04} (2004) 048, [\href{http://arxiv.org/abs/hep-th/0401129}{{\tt hep-th/0401129}}].

\bibitem{Chong:2005hr}
Z.~W. Chong, M.~Cvetic, H.~Lu, and C.~N. Pope, {\it {General non-extremal rotating black holes in minimal five-dimensional gauged supergravity}},  {\em Phys. Rev. Lett.} {\bf 95} (2005) 161301, [\href{http://arxiv.org/abs/hep-th/0506029}{{\tt hep-th/0506029}}].

\bibitem{Kunduri:2006ek}
H.~K. Kunduri, J.~Lucietti, and H.~S. Reall, {\it {Supersymmetric multi-charge AdS(5) black holes}},  {\em JHEP} {\bf 04} (2006) 036, [\href{http://arxiv.org/abs/hep-th/0601156}{{\tt hep-th/0601156}}].

\bibitem{Cabo-Bizet:2018ehj}
A.~Cabo-Bizet, D.~Cassani, D.~Martelli, and S.~Murthy, {\it {Microscopic origin of the Bekenstein-Hawking entropy of supersymmetric AdS$_{5}$ black holes}},  {\em JHEP} {\bf 10} (2019) 062, [\href{http://arxiv.org/abs/1810.11442}{{\tt arXiv:1810.11442}}].

\bibitem{Choi:2018hmj}
S.~Choi, J.~Kim, S.~Kim, and J.~Nahmgoong, {\it {Large AdS black holes from QFT}},  \href{http://arxiv.org/abs/1810.12067}{{\tt arXiv:1810.12067}}.

\bibitem{Benini:2018ywd}
F.~Benini and E.~Milan, {\it {Black Holes in 4D $\mathcal{N}$=4 Super-Yang-Mills Field Theory}},  {\em Phys. Rev. X} {\bf 10} (2020), no.~2 021037, [\href{http://arxiv.org/abs/1812.09613}{{\tt arXiv:1812.09613}}].

\bibitem{Choi:2018vbz}
S.~Choi, J.~Kim, S.~Kim, and J.~Nahmgoong, {\it {Comments on deconfinement in AdS/CFT}},  \href{http://arxiv.org/abs/1811.08646}{{\tt arXiv:1811.08646}}.

\bibitem{Honda:2019cio}
M.~Honda, {\it {Quantum Black Hole Entropy from 4d Supersymmetric Cardy formula}},  {\em Phys. Rev. D} {\bf 100} (2019), no.~2 026008, [\href{http://arxiv.org/abs/1901.08091}{{\tt arXiv:1901.08091}}].

\bibitem{ArabiArdehali:2019tdm}
A.~Arabi~Ardehali, {\it {Cardy-like asymptotics of the 4d $ \mathcal{N}=4 $ index and AdS$_{5}$ blackholes}},  {\em JHEP} {\bf 06} (2019) 134, [\href{http://arxiv.org/abs/1902.06619}{{\tt arXiv:1902.06619}}].

\bibitem{Kim:2019yrz}
J.~Kim, S.~Kim, and J.~Song, {\it {A 4d $ \mathcal{N} $ = 1 Cardy Formula}},  {\em JHEP} {\bf 01} (2021) 025, [\href{http://arxiv.org/abs/1904.03455}{{\tt arXiv:1904.03455}}].

\bibitem{Cabo-Bizet:2019osg}
A.~Cabo-Bizet, D.~Cassani, D.~Martelli, and S.~Murthy, {\it {The asymptotic growth of states of the 4d $ \mathcal{N}=1 $ superconformal index}},  {\em JHEP} {\bf 08} (2019) 120, [\href{http://arxiv.org/abs/1904.05865}{{\tt arXiv:1904.05865}}].

\bibitem{Amariti:2019mgp}
A.~Amariti, I.~Garozzo, and G.~Lo~Monaco, {\it {Entropy function from toric geometry}},  {\em Nucl. Phys. B} {\bf 973} (2021) 115571, [\href{http://arxiv.org/abs/1904.10009}{{\tt arXiv:1904.10009}}].

\bibitem{Goldstein:2020yvj}
K.~Goldstein, V.~Jejjala, Y.~Lei, S.~van Leuven, and W.~Li, {\it {Residues, modularity, and the Cardy limit of the 4d $ \mathcal{N} $ = 4 superconformal index}},  {\em JHEP} {\bf 04} (2021) 216, [\href{http://arxiv.org/abs/2011.06605}{{\tt arXiv:2011.06605}}].

\bibitem{Jejjala:2021hlt}
V.~Jejjala, Y.~Lei, S.~van Leuven, and W.~Li, {\it {SL(3, \ensuremath{\mathbb{Z}}) Modularity and New Cardy limits of the $ \mathcal{N} $ = 4 superconformal index}},  {\em JHEP} {\bf 11} (2021) 047, [\href{http://arxiv.org/abs/2104.07030}{{\tt arXiv:2104.07030}}].

\bibitem{Closset:2017bse}
C.~Closset, H.~Kim, and B.~Willett, {\it {$ \mathcal{N} $ = 1 supersymmetric indices and the four-dimensional A-model}},  {\em JHEP} {\bf 08} (2017) 090, [\href{http://arxiv.org/abs/1707.05774}{{\tt arXiv:1707.05774}}].

\bibitem{Benini:2018mlo}
F.~Benini and E.~Milan, {\it {A Bethe Ansatz type formula for the superconformal index}},  {\em Commun. Math. Phys.} {\bf 376} (2020), no.~2 1413--1440, [\href{http://arxiv.org/abs/1811.04107}{{\tt arXiv:1811.04107}}].

\bibitem{Lanir:2019abx}
A.~Lanir, A.~Nedelin, and O.~Sela, {\it {Black hole entropy function for toric theories via Bethe Ansatz}},  {\em JHEP} {\bf 04} (2020) 091, [\href{http://arxiv.org/abs/1908.01737}{{\tt arXiv:1908.01737}}].

\bibitem{GonzalezLezcano:2019nca}
A.~Gonz\'alez~Lezcano and L.~A. Pando~Zayas, {\it {Microstate counting via Bethe Ans\"atze in the 4d $ \mathcal{N} $ = 1 superconformal index}},  {\em JHEP} {\bf 03} (2020) 088, [\href{http://arxiv.org/abs/1907.12841}{{\tt arXiv:1907.12841}}].

\bibitem{Benini:2020gjh}
F.~Benini, E.~Colombo, S.~Soltani, A.~Zaffaroni, and Z.~Zhang, {\it {Superconformal indices at large $N$ and the entropy of AdS$_5$ $\times$ SE$_5$ black holes}},  {\em Class. Quant. Grav.} {\bf 37} (2020), no.~21 215021, [\href{http://arxiv.org/abs/2005.12308}{{\tt arXiv:2005.12308}}].

\bibitem{Mamroud:2022msu}
O.~Mamroud, {\it {The SUSY index beyond the Cardy limit}},  {\em JHEP} {\bf 01} (2024) 111, [\href{http://arxiv.org/abs/2212.11925}{{\tt arXiv:2212.11925}}].

\bibitem{Aharony:2024ntg}
O.~Aharony, O.~Mamroud, S.~Nowik, and M.~Weissman, {\it {The Bethe Ansatz for the superconformal index with unequal angular momenta}},  \href{http://arxiv.org/abs/2402.03977}{{\tt arXiv:2402.03977}}.

\bibitem{Cabo-Bizet:2019eaf}
A.~Cabo-Bizet and S.~Murthy, {\it {Supersymmetric phases of 4d $ \mathcal{N} $ = 4 SYM at large $N$}},  {\em JHEP} {\bf 09} (2020) 184, [\href{http://arxiv.org/abs/1909.09597}{{\tt arXiv:1909.09597}}].

\bibitem{Cabo-Bizet:2020nkr}
A.~Cabo-Bizet, D.~Cassani, D.~Martelli, and S.~Murthy, {\it {The large-$N$ limit of the 4d $ \mathcal{N} $ = 1 superconformal index}},  {\em JHEP} {\bf 11} (2020) 150, [\href{http://arxiv.org/abs/2005.10654}{{\tt arXiv:2005.10654}}].

\bibitem{Choi:2021rxi}
S.~Choi, S.~Jeong, S.~Kim, and E.~Lee, {\it {Exact QFT duals of AdS black holes}},  \href{http://arxiv.org/abs/2111.10720}{{\tt arXiv:2111.10720}}.

\bibitem{Choi:2023tiq}
S.~Choi, S.~Kim, and J.~Song, {\it {Large $N$ Universality of 4d $\mathcal{N}=1$ Superconformal Index and AdS Black Holes}},  \href{http://arxiv.org/abs/2309.07614}{{\tt arXiv:2309.07614}}.

\bibitem{Gaiotto:2021xce}
D.~Gaiotto and J.~H. Lee, {\it {The Giant Graviton Expansion}},  \href{http://arxiv.org/abs/2109.02545}{{\tt arXiv:2109.02545}}.

\bibitem{Imamura:2021ytr}
Y.~Imamura, {\it {Finite-N superconformal index via the AdS/CFT correspondence}},  {\em PTEP} {\bf 2021} (2021), no.~12 123B05, [\href{http://arxiv.org/abs/2108.12090}{{\tt arXiv:2108.12090}}].

\bibitem{Lee:2022vig}
J.~H. Lee, {\it {Exact stringy microstates from gauge theories}},  {\em JHEP} {\bf 11} (2022) 137, [\href{http://arxiv.org/abs/2204.09286}{{\tt arXiv:2204.09286}}].

\bibitem{Murthy:2022ien}
S.~Murthy, {\it {Unitary matrix models, free fermions, and the giant graviton expansion}},  {\em Pure Appl. Math. Quart.} {\bf 19} (2023), no.~1 299--340, [\href{http://arxiv.org/abs/2202.06897}{{\tt arXiv:2202.06897}}].

\bibitem{Liu:2022olj}
J.~T. Liu and N.~J. Rajappa, {\it {Finite N indices and the giant graviton expansion}},  {\em JHEP} {\bf 04} (2023) 078, [\href{http://arxiv.org/abs/2212.05408}{{\tt arXiv:2212.05408}}].

\bibitem{Lin:2022gbu}
H.~Lin, {\it {Coherent state operators, giant gravitons, and gauge-gravity correspondence}},  {\em Annals Phys.} {\bf 451} (2023) 169248, [\href{http://arxiv.org/abs/2212.14002}{{\tt arXiv:2212.14002}}].

\bibitem{Eniceicu:2023uvd}
D.~S. Eniceicu, {\it {Comments on the Giant-Graviton Expansion of the Superconformal Index}},  \href{http://arxiv.org/abs/2302.04887}{{\tt arXiv:2302.04887}}.

\bibitem{Lee:2023iil}
J.~H. Lee, {\it {Trace relations and open string vacua}},  \href{http://arxiv.org/abs/2312.00242}{{\tt arXiv:2312.00242}}.

\bibitem{Eleftheriou:2023jxr}
G.~Eleftheriou, S.~Murthy, and M.~Rossell\'o, {\it {The giant graviton expansion in $AdS_5 \times S^5$}},  \href{http://arxiv.org/abs/2312.14921}{{\tt arXiv:2312.14921}}.

\bibitem{McGreevy:2000cw}
J.~McGreevy, L.~Susskind, and N.~Toumbas, {\it {Invasion of the giant gravitons from Anti-de Sitter space}},  {\em JHEP} {\bf 06} (2000) 008, [\href{http://arxiv.org/abs/hep-th/0003075}{{\tt hep-th/0003075}}].

\bibitem{Grisaru:2000zn}
M.~T. Grisaru, R.~C. Myers, and O.~Tafjord, {\it {SUSY and goliath}},  {\em JHEP} {\bf 08} (2000) 040, [\href{http://arxiv.org/abs/hep-th/0008015}{{\tt hep-th/0008015}}].

\bibitem{Hashimoto:2000zp}
A.~Hashimoto, S.~Hirano, and N.~Itzhaki, {\it {Large branes in AdS and their field theory dual}},  {\em JHEP} {\bf 08} (2000) 051, [\href{http://arxiv.org/abs/hep-th/0008016}{{\tt hep-th/0008016}}].

\bibitem{Balasubramanian:2001nh}
V.~Balasubramanian, M.~Berkooz, A.~Naqvi, and M.~J. Strassler, {\it {Giant gravitons in conformal field theory}},  {\em JHEP} {\bf 04} (2002) 034, [\href{http://arxiv.org/abs/hep-th/0107119}{{\tt hep-th/0107119}}].

\bibitem{Imamura:2021dya}
Y.~Imamura and S.~Murayama, {\it {Holographic index calculation for Argyres\textendash{}Douglas and Minahan\textendash{}Nemeschansky theories}},  {\em PTEP} {\bf 2022} (2022), no.~11 113B01, [\href{http://arxiv.org/abs/2110.14897}{{\tt arXiv:2110.14897}}].

\bibitem{Fujiwara:2021xgu}
S.~Fujiwara, Y.~Imamura, and T.~Mori, {\it {Flavor symmetries of six-dimensional ${\cal N}=(1,0)$ theories from AdS/CFT correspondence}},  {\em JHEP} {\bf 05} (2021) 221, [\href{http://arxiv.org/abs/2103.16094}{{\tt arXiv:2103.16094}}].

\bibitem{Arai:2020uwd}
R.~Arai, S.~Fujiwara, Y.~Imamura, T.~Mori, and D.~Yokoyama, {\it {Finite-$N$ corrections to the M-brane indices}},  {\em JHEP} {\bf 11} (2020) 093, [\href{http://arxiv.org/abs/2007.05213}{{\tt arXiv:2007.05213}}].

\bibitem{Arai:2020qaj}
R.~Arai, S.~Fujiwara, Y.~Imamura, and T.~Mori, {\it {Schur index of the ${\cal N}=4$ $U(N)$ supersymmetric Yang-Mills theory via the AdS/CFT correspondence}},  {\em Phys. Rev. D} {\bf 101} (2020), no.~8 086017, [\href{http://arxiv.org/abs/2001.11667}{{\tt arXiv:2001.11667}}].

\bibitem{Arai:2019aou}
R.~Arai, S.~Fujiwara, Y.~Imamura, and T.~Mori, {\it {Finite $N$ corrections to the superconformal index of toric quiver gauge theories}},  {\em PTEP} {\bf 2020} (2020), no.~4 043B09, [\href{http://arxiv.org/abs/1911.10794}{{\tt arXiv:1911.10794}}].

\bibitem{Arai:2019wgv}
R.~Arai, S.~Fujiwara, Y.~Imamura, and T.~Mori, {\it {Finite $N$ corrections to the superconformal index of orbifold quiver gauge theories}},  {\em JHEP} {\bf 10} (2019) 243, [\href{http://arxiv.org/abs/1907.05660}{{\tt arXiv:1907.05660}}].

\bibitem{Arai:2019xmp}
R.~Arai and Y.~Imamura, {\it {Finite $N$ Corrections to the Superconformal Index of S-fold Theories}},  {\em PTEP} {\bf 2019} (2019), no.~8 083B04, [\href{http://arxiv.org/abs/1904.09776}{{\tt arXiv:1904.09776}}].

\bibitem{Choi:2022ovw}
S.~Choi, S.~Kim, E.~Lee, and J.~Lee, {\it {From giant gravitons to black holes}},  {\em JHEP} {\bf 11} (2023) 086, [\href{http://arxiv.org/abs/2207.05172}{{\tt arXiv:2207.05172}}].

\bibitem{Kinney:2005ej}
J.~Kinney, J.~M. Maldacena, S.~Minwalla, and S.~Raju, {\it {An Index for 4 dimensional super conformal theories}},  {\em Commun. Math. Phys.} {\bf 275} (2007) 209--254, [\href{http://arxiv.org/abs/hep-th/0510251}{{\tt hep-th/0510251}}].

\bibitem{Hagedorn:1965st}
R.~Hagedorn, {\it {Statistical thermodynamics of strong interactions at high-energies}},  {\em Nuovo Cim. Suppl.} {\bf 3} (1965) 147--186.

\bibitem{Atick:1988si}
J.~J. Atick and E.~Witten, {\it {The Hagedorn Transition and the Number of Degrees of Freedom of String Theory}},  {\em Nucl. Phys. B} {\bf 310} (1988) 291--334.

\bibitem{Beccaria:2023hip}
M.~Beccaria and A.~Cabo-Bizet, {\it {Large black hole entropy from the giant brane expansion}},  \href{http://arxiv.org/abs/2308.05191}{{\tt arXiv:2308.05191}}.

\bibitem{Imamura:2022aua}
Y.~Imamura, {\it {Analytic continuation for giant gravitons}},  {\em PTEP} {\bf 2022} (2022), no.~10 103B02, [\href{http://arxiv.org/abs/2205.14615}{{\tt arXiv:2205.14615}}].

\bibitem{Fujiwara:2023bdc}
S.~Fujiwara, Y.~Imamura, T.~Mori, S.~Murayama, and D.~Yokoyama, {\it {Simple-Sum Giant Graviton Expansions for Orbifolds and Orientifolds}},  \href{http://arxiv.org/abs/2310.03332}{{\tt arXiv:2310.03332}}.

\bibitem{Aharony:2021zkr}
O.~Aharony, F.~Benini, O.~Mamroud, and E.~Milan, {\it {A gravity interpretation for the Bethe Ansatz expansion of the $\mathcal{N}=4$ SYM index}},  {\em Phys. Rev. D} {\bf 104} (2021) 086026, [\href{http://arxiv.org/abs/2104.13932}{{\tt arXiv:2104.13932}}].

\bibitem{Gukov:2006jk}
S.~Gukov and E.~Witten, {\it {Gauge Theory, Ramification, And The Geometric Langlands Program}},  \href{http://arxiv.org/abs/hep-th/0612073}{{\tt hep-th/0612073}}.

\bibitem{Nakayama:2011pa}
Y.~Nakayama, {\it {4D and 2D superconformal index with surface operator}},  {\em JHEP} {\bf 08} (2011) 084, [\href{http://arxiv.org/abs/1105.4883}{{\tt arXiv:1105.4883}}].

\bibitem{Chen:2023lzq}
Y.~Chen, M.~Heydeman, Y.~Wang, and M.~Zhang, {\it {Probing Supersymmetric Black Holes with Surface Defects}},  \href{http://arxiv.org/abs/2306.05463}{{\tt arXiv:2306.05463}}.

\bibitem{Cabo-Bizet:2023ejm}
A.~Cabo-Bizet, M.~David, and A.~Gonz\'alez~Lezcano, {\it {Thermodynamics of black holes with probe D-branes}},  \href{http://arxiv.org/abs/2312.12533}{{\tt arXiv:2312.12533}}.

\bibitem{Romelsberger:2005eg}
C.~Romelsberger, {\it {Counting chiral primaries in N = 1, d=4 superconformal field theories}},  {\em Nucl. Phys. B} {\bf 747} (2006) 329--353, [\href{http://arxiv.org/abs/hep-th/0510060}{{\tt hep-th/0510060}}].

\bibitem{Cassani:2021fyv}
D.~Cassani and Z.~Komargodski, {\it {EFT and the SUSY Index on the 2nd Sheet}},  {\em SciPost Phys.} {\bf 11} (2021) 004, [\href{http://arxiv.org/abs/2104.01464}{{\tt arXiv:2104.01464}}].

\bibitem{Felder_2000}
G.~Felder and A.~Varchenko, {\it The elliptic gamma function and {$SL(3,~Z)\ltimes Z3$}},  {\em Advances in Mathematics} {\bf 156} (dec, 2000) 44--76.

\bibitem{Hosseini:2017mds}
S.~M. Hosseini, K.~Hristov, and A.~Zaffaroni, {\it {An extremization principle for the entropy of rotating BPS black holes in AdS$_{5}$}},  {\em JHEP} {\bf 07} (2017) 106, [\href{http://arxiv.org/abs/1705.05383}{{\tt arXiv:1705.05383}}].

\bibitem{Frolov:1989jh}
V.~P. Frolov and K.~S. Thorne, {\it {Renormalized Stress - Energy Tensor Near the Horizon of a Slowly Evolving, Rotating Black Hole}},  {\em Phys. Rev. D} {\bf 39} (1989) 2125--2154.

\bibitem{Constable:2002xt}
N.~R. Constable, J.~Erdmenger, Z.~Guralnik, and I.~Kirsch, {\it {Intersecting D-3 branes and holography}},  {\em Phys. Rev. D} {\bf 68} (2003) 106007, [\href{http://arxiv.org/abs/hep-th/0211222}{{\tt hep-th/0211222}}].

\bibitem{Kachru:1998ys}
S.~Kachru and E.~Silverstein, {\it {4-D conformal theories and strings on orbifolds}},  {\em Phys. Rev. Lett.} {\bf 80} (1998) 4855--4858, [\href{http://arxiv.org/abs/hep-th/9802183}{{\tt hep-th/9802183}}].

\bibitem{Lawrence:1998ja}
A.~E. Lawrence, N.~Nekrasov, and C.~Vafa, {\it {On conformal field theories in four-dimensions}},  {\em Nucl. Phys. B} {\bf 533} (1998) 199--209, [\href{http://arxiv.org/abs/hep-th/9803015}{{\tt hep-th/9803015}}].

\bibitem{Nakayama:2005mf}
Y.~Nakayama, {\it {Index for orbifold quiver gauge theories}},  {\em Phys. Lett. B} {\bf 636} (2006) 132--136, [\href{http://arxiv.org/abs/hep-th/0512280}{{\tt hep-th/0512280}}].

\bibitem{Martelli:2004wu}
D.~Martelli and J.~Sparks, {\it {Toric geometry, Sasaki-Einstein manifolds and a new infinite class of AdS/CFT duals}},  {\em Commun. Math. Phys.} {\bf 262} (2006) 51--89, [\href{http://arxiv.org/abs/hep-th/0411238}{{\tt hep-th/0411238}}].

\bibitem{Benvenuti:2004dy}
S.~Benvenuti, S.~Franco, A.~Hanany, D.~Martelli, and J.~Sparks, {\it {An Infinite family of superconformal quiver gauge theories with Sasaki-Einstein duals}},  {\em JHEP} {\bf 06} (2005) 064, [\href{http://arxiv.org/abs/hep-th/0411264}{{\tt hep-th/0411264}}].

\bibitem{Klebanov:1998hh}
I.~R. Klebanov and E.~Witten, {\it {Superconformal field theory on three-branes at a Calabi-Yau singularity}},  {\em Nucl. Phys. B} {\bf 536} (1998) 199--218, [\href{http://arxiv.org/abs/hep-th/9807080}{{\tt hep-th/9807080}}].

\bibitem{Hanany:2005ve}
A.~Hanany and K.~D. Kennaway, {\it {Dimer models and toric diagrams}},  \href{http://arxiv.org/abs/hep-th/0503149}{{\tt hep-th/0503149}}.

\bibitem{Franco:2005sm}
S.~Franco, A.~Hanany, D.~Martelli, J.~Sparks, D.~Vegh, and B.~Wecht, {\it {Gauge theories from toric geometry and brane tilings}},  {\em JHEP} {\bf 01} (2006) 128, [\href{http://arxiv.org/abs/hep-th/0505211}{{\tt hep-th/0505211}}].

\bibitem{Franco:2005rj}
S.~Franco, A.~Hanany, K.~D. Kennaway, D.~Vegh, and B.~Wecht, {\it {Brane dimers and quiver gauge theories}},  {\em JHEP} {\bf 01} (2006) 096, [\href{http://arxiv.org/abs/hep-th/0504110}{{\tt hep-th/0504110}}].

\bibitem{Feng:2005gw}
B.~Feng, Y.-H. He, K.~D. Kennaway, and C.~Vafa, {\it {Dimer models from mirror symmetry and quivering amoebae}},  {\em Adv. Theor. Math. Phys.} {\bf 12} (2008), no.~3 489--545, [\href{http://arxiv.org/abs/hep-th/0511287}{{\tt hep-th/0511287}}].

\bibitem{Intriligator:2003jj}
K.~A. Intriligator and B.~Wecht, {\it {The Exact superconformal R symmetry maximizes a}},  {\em Nucl. Phys. B} {\bf 667} (2003) 183--200, [\href{http://arxiv.org/abs/hep-th/0304128}{{\tt hep-th/0304128}}].

\bibitem{Martelli:2005tp}
D.~Martelli, J.~Sparks, and S.-T. Yau, {\it {The Geometric dual of a-maximisation for Toric Sasaki-Einstein manifolds}},  {\em Commun. Math. Phys.} {\bf 268} (2006) 39--65, [\href{http://arxiv.org/abs/hep-th/0503183}{{\tt hep-th/0503183}}].

\bibitem{Guica:2008mu}
M.~Guica, T.~Hartman, W.~Song, and A.~Strominger, {\it {The Kerr/CFT Correspondence}},  {\em Phys. Rev. D} {\bf 80} (2009) 124008, [\href{http://arxiv.org/abs/0809.4266}{{\tt arXiv:0809.4266}}].

\bibitem{Chow:2008dp}
D.~D.~K. Chow, M.~Cvetic, H.~Lu, and C.~N. Pope, {\it {Extremal Black Hole/CFT Correspondence in (Gauged) Supergravities}},  {\em Phys. Rev. D} {\bf 79} (2009) 084018, [\href{http://arxiv.org/abs/0812.2918}{{\tt arXiv:0812.2918}}].

\bibitem{Lin:2004nb}
H.~Lin, O.~Lunin, and J.~M. Maldacena, {\it {Bubbling AdS space and 1/2 BPS geometries}},  {\em JHEP} {\bf 10} (2004) 025, [\href{http://arxiv.org/abs/hep-th/0409174}{{\tt hep-th/0409174}}].

\bibitem{Gomis:2007fi}
J.~Gomis and S.~Matsuura, {\it {Bubbling surface operators and S-duality}},  {\em JHEP} {\bf 06} (2007) 025, [\href{http://arxiv.org/abs/0704.1657}{{\tt arXiv:0704.1657}}].

\bibitem{Bah:2021hei}
I.~Bah, F.~Bonetti, R.~Minasian, and E.~Nardoni, {\it {M5-brane sources, holography, and Argyres-Douglas theories}},  {\em JHEP} {\bf 11} (2021) 140, [\href{http://arxiv.org/abs/2106.01322}{{\tt arXiv:2106.01322}}].

\bibitem{Bah:2021mzw}
I.~Bah, F.~Bonetti, R.~Minasian, and E.~Nardoni, {\it {Holographic Duals of Argyres-Douglas Theories}},  {\em Phys. Rev. Lett.} {\bf 127} (2021), no.~21 211601, [\href{http://arxiv.org/abs/2105.11567}{{\tt arXiv:2105.11567}}].

\bibitem{Chang:2022mjp}
C.-M. Chang and Y.-H. Lin, {\it {Words to describe a black hole}},  {\em JHEP} {\bf 02} (2023) 109, [\href{http://arxiv.org/abs/2209.06728}{{\tt arXiv:2209.06728}}].

\bibitem{Choi:2022caq}
S.~Choi, S.~Kim, E.~Lee, and J.~Park, {\it {The shape of non-graviton operators for $SU(2)$}},  \href{http://arxiv.org/abs/2209.12696}{{\tt arXiv:2209.12696}}.

\bibitem{Choi:2023znd}
S.~Choi, S.~Kim, E.~Lee, S.~Lee, and J.~Park, {\it {Towards quantum black hole microstates}},  \href{http://arxiv.org/abs/2304.10155}{{\tt arXiv:2304.10155}}.

\bibitem{Choi:2023vdm}
J.~Choi, S.~Choi, S.~Kim, J.~Lee, and S.~Lee, {\it {Finite $N$ black hole cohomologies}},  \href{http://arxiv.org/abs/2312.16443}{{\tt arXiv:2312.16443}}.

\bibitem{Chang:2023zqk}
C.-M. Chang, L.~Feng, Y.-H. Lin, and Y.-X. Tao, {\it {Decoding stringy near-supersymmetric black holes}},  \href{http://arxiv.org/abs/2306.04673}{{\tt arXiv:2306.04673}}.

\bibitem{Budzik:2023vtr}
K.~Budzik, H.~Murali, and P.~Vieira, {\it {Following Black Hole States}},  \href{http://arxiv.org/abs/2306.04693}{{\tt arXiv:2306.04693}}.

\end{thebibliography}\endgroup

\end{document}